\documentclass[12pt]{article}
\usepackage{epsfig,amsfonts}
\usepackage{amsmath,amssymb}

\newcommand{\be}{\begin{equation}}
\newcommand{\ee}{\end{equation}}
\newcommand{\bea}{\begin{eqnarray}}
\newcommand{\eea}{\end{eqnarray}}
\newcommand{\nonu}{\nonumber\\}
\def\d{\mathrm{d}}
\def\eq#1{(\ref{#1})}
\def\hf{\frac{1}{2}}
\def\ord{{\cal O}}

\begin{document}

\title{Wegner-Houghton equation in low dimensions}
\author{J.M. Carmona$^a$, J. Polonyi$^b$, A. Taranc\'on$^c$\\[0.5em]
$^a$ {\small Dipartimento di Fisica,}
{\small Universit\`a di Pisa,}\\
{\small Via Buonarroti, 2, Ed. B, 56127 Pisa (Italy)}\\[0.3em]
$^b$ {\small Laboratoire de Physique Th\'eorique,}
{\small Universit\'e Louis Pasteur,}\\
{\small 3, rue de l'Universit\'e, 67084 Strasbourg Cedex (France), and}\\
{\small Department of Atomic Physics, L. E\"otv\"os University,
Budapest, Hungary}\\[0.3em]
$^c$ {\small Departamento de F\'{\i}sica Te\'orica,}
{\small Universidad de Zaragoza,}\\
{\small Pedro Cerbuna 12, 50009 Zaragoza (Spain)}\\[0.3em]
{\small \tt carmona@mailbox.difiunipi.it}\\
{\small \tt polonyi@fresnel.u-strasbg.fr}\\
{\small \tt tarancon@sol.unizar.es}\\[0.5em]}

\maketitle

\begin{abstract}
We consider scalar field theories in dimensions lower than four in the
context of the Wegner-Houghton renormalization group equations (WHRG).
The renormalized trajectory makes
a non-perturbative interpolation
between the ultraviolet and the infrared scaling regimes.
Strong indication is found that in two dimensions and below
the models with polynomial interaction
are always non-perturbative in the infrared scaling regime.
Finally we check that these results do not depend on the regularization
and we develop a lattice version of the WHRG in two dimensions.
\end{abstract}

\vskip 2 mm

\vfill

Renormalization group.
Low dimensional theories.
Lattice renormalization group.

\thispagestyle{empty}

\newpage

\section{Introduction}
\label{sect1}
Asymptotically free models gained importance both in High Energy
and in Condensed Matter Physics. This is obvious for the latter,
where the low dimensional phenomena invoke effective theories
below the upper critical dimension, $d=4$, which are super-renormalizable
and their Hamiltonians contain asymptotically free coupling constants only.
In the former case the the asymptotically free models can provide
structureless high energy physics where the cutoff can be pushed
away at will. What kind of non-perturbative mechanisms do we encounter as
we follow the renormalization group flow of an asymptotically
free model? To distinguish different
finite energy mechanisms we recall that unless the model
possesses unbroken scale invariance there are always
two scaling regimes, an ultraviolet and an infrared one
separated by a crossover at the characteristic scale of the model.
If there is a dimensional coupling constant, say a mass parameter
$m$, in the Lagrangian then the UV and IR scaling regimes
are separated by a crossover at $p=p_{cr}\approx m$ because
$m^2/p^2$ or $p^2/m^2$ is treated as a small quantity in the
UV or the IR side, respectively. The IR scaling laws are
trivial because the radiative corrections to the
evolution are suppressed by $p^2/m^2$ and what is left
is the scale dependence governed by the canonical dimensions
of the coupling constants. If there are no dimensional
coupling constants in the Lagrangian, i.e. the model
possesses classical scale invariance then a crossover
can be identified where one of the coupling constants reach
the value 1, $g(p_{cr})=1$. The expression for the dimensionless
running coupling constant $g(p)$ must contain a dimensional
parameter, usually called the $\Lambda$-parameter and
$p_{cr}\approx\Lambda_{\phi^4}$.\footnote{It is worthwhile noting that one can
introduce the $\Lambda$-parameter even in the presence of
any parameter with positive mass dimension in the Lagrangian
so long as the IR dynamics remains stable when this latter
is removed. In fact, let us write this parameter as $m^\ell$
where $m$ defined in this manner is the mass scale of the
model. The running coupling
constant depends on the ratio $m/p$, $g(p)=f(m/p)$ where
the limit $m\to0$ is convergent. In other words, the
evolution of the classically scale invariant system is recovered
as $p\to\infty$.}

Non-perturbative effects may originate in the IR
scaling regimes. The classically scale invariant models
at the lower critical dimension, such as the two-dimensional
sigma and the Gross-Neveu model do not support long range order.
Thus the IR scaling laws must contain a dynamically generated
mass scale, an effect generated by
the infrared or collinear divergences.

The asymptotically free coupling constants may generate
non-perturbative effects in the UV scaling regime, as well,
due to their growth. The question is whether the scale
parameter $\Lambda_{\phi^4}$
or the mass $m$ is reached first as we lower the cutoff.
The system remains perturbative for $m>\Lambda_{\phi^4}$ because
the IR scaling laws cut off the growth of the asymptotically
free coupling constants before they would reach a dangerously
large value. The dynamics of the modes $p<\Lambda_{\phi^4}$
becomes non-perturbative when $\Lambda_{\phi^4}>m$.
The four-dimensional Yang-Mills
models develop linearly rising potential
between static external charges. This is believed to
happen due to the large value of the asymptotically free
coupling constant at the ultraviolet side of the
crossover, indicated by the infrared Landau
pole in perturbative QCD. The quark masses are supposed to
play no important role in the vacuum structure and the
chiral limit $m\to0$ is assumed to be safe and convergent,
though non-perturbative. 

The infrared singularities are easier to isolate than the non-perturbative
effects arising at the IR end of the UV scaling regime. It
has already been achieved for non-asymptotically free
models \cite{irqed} and superrenormalizable theories
\cite{jate}. The more complex non-perturbative features should arise
at the ultraviolet side of the crossover where no asymptotic
analysis is available. 

The goal of the present paper is the study of the scaling laws 
and their consequences at the IR end of the UV scaling regime.
According to our best knowledge this source of the non-perturbative
effects in asymptotically free theories 
has not been studied before in a systematic manner.
The problem is rather involved and demanding, we restrict
ourselves to outline only the way the more detailed analysis 
should be done.
In order to defuse the infrared problem we turn to the scalar model with
polynomial interaction in the phase where no spontaneous
symmetry breaking occurs. The term $\phi^n$ in the Lagrangian
of the $d$-dimensional scalar model is asymptotically
free for $n<2d/(d-2)$, so we have to trace the evolution
of a large number of coupling constants at low
dimensions. We present numerical evidences that
the polinomial interactions at the lower critical dimension
or below where infinitely many operators are relevant in the
UV scaling regime always become strong, non-perturbative.

The study of the asymptotically free three-dimensional $\phi^6$ model
can bring some light into the IR Landau pole problem of four-dimensional
gauge theories by considering the role of the non-renormalizable
couplings in the IR extrapolation of the UV scaling regime.
We will show that they affect considerably the evolution of the
relevant couplings. This points to the possibility that the
non-renormalizable couplings could suppress the Landau pole. The
full solution of the theory where all possible
coupling constants are followed
by the renormalization group equation should necessarily yield
a non-singular renormalization group flow so long as the theory
possesses local interactions. This suggests that the introduction of
the hadronic composite operators in QCD could defuse its
Landau pole problem \cite{polonyi95}.

Another issue addressed in this paper is the manner in which the
influence of the non-renormalizable operators on the dynamics
is suppressed and the universal physics is reached
as the cutoff is lowered. The usual argument based on the
linearization of the blocking relation is not
obviously applicable due to the presence of infinitely many 
relevant operators. This question is discussed
in the framework of lattice regulation by
tailoring the Wegner-Houghton equation for lattice
regulated models. As a result, the approach to the
low energy physics can be compared for the momentum space
cutoff regularization in the continuum and for the lattice regularization,
and the universality, the regulator independence, 
can be established well beyond the linearized approximation
of the blocking relations.

The organization of this paper is the following. The infinitesimal
renormalization group step, the Wegner-Houghton equation, is
described in Section II. The running coupling constants are
introduced and their evolution equations are given in Section III
for scalar models. The asymptotic UV evolution
is discussed in Section IV for $d=2,~3$ and 4. Section V is devoted
to the demonstration of the difficulties in finding a perturbative
model in $d=2$.
The Wegner-Houghton equation is derived in the lattice regularization
and its solution is presented in Section VI. Section VII is for
the conclusions.

\section{The Wegner-Houghton equation}
There are different ways the mixing of a large number of
operators can be traced down. The Wegner-Houghton 
equation \cite{wh73},
which we use in the local potential approximation in this work,
is the simplest implementation of
the Kadanoff-Wilson blocking \cite{wrg} in the momentum space
and produces the cutoff dependence of the bare action along
the renormalized trajectory. Other methods work with the 
effective action where the infrared cutoff dependence is
sought \cite{others}. Different schemes should agree in the
infrared limit where few long wavelength modes are left only
in the system. We shall make two approximations in computing
the blocked action, the truncation of the gradient expansion
at the leading order, the local potential approximation \cite{locp},
and the truncation of the Taylor expansion of the local potential 
in the field variable \cite{trpot}.
The higher order terms of the gradient expansion are 
non-renormalizable according to the power counting.
We believe that these coupling constants which are irrelevant
in the ultraviolet scaling regime do not modify our qualitative
conclusion.

The Wegner-Houghton (WH) equation~\cite{wh73} describes the evolution
of the effective action as the cutoff is lowered. As mentioned in the
Introduction we shall consider a scalar model with an intrinsic mass
scale which allows to clearly distinguish the UV and IR scaling regimes. 
We derive the WH equation for a scalar field theory by using
a sharp momentum space cutoff~\cite{polonyi95,polonyi97}:
we will call this regularization procedure the \textit{continuum
regularization.} In Section~\ref{lattice} we consider an
alternative, the \textit{lattice regularization.}

Denote the bare action by $S_k[\phi]$, where $k$ is the UV cutoff. Then,
according to the usual Wilson-Kadanoff procedure,
\begin{equation}
e^{-\frac{1}{\hbar}S_{k'}[\phi]}=\int\mathcal{D}\phi'\,
e^{-\frac{1}{\hbar}S_{k}[\phi+\phi']}
\label{blocking}
\end{equation}
where $k'<k$ in the Euclidean space-time. The Fourier transform of the
fields $\phi(x)$ and $\phi'(x)$ is non-vanishing only for $p<k'$ and
$k'<p<k$, respectively. The right hand side is
evaluated by means of the loop expansion, so that Eq.~(\ref{blocking})
gives
\begin{equation}
S_{k'}[\phi]=S_k[\phi+\phi'_0]+\frac{\hbar}{2}
\mathrm{tr}\log\delta^2 S+\ord(\hbar^2),
\label{loopexp}
\end{equation}
where
\begin{equation}
\delta^2 S(x,y)=\frac{\delta^2 S_k[\phi+\phi'_0]}{\delta\phi'(x)
\delta\phi'(y)}\, ,
\end{equation}
and the saddle point, $\phi'_0$, is defined by the extremum condition
\begin{equation}
\frac{\delta S_k[\phi+\phi'_0]}{\delta\phi'}=0,
\end{equation}
in which the infrared background field, $\phi(x)$, is held fixed.
It can be proved that the saddle point is trivial, $\phi'_0=0$, as long as
the matrix $\delta^2S(x,y)$ is invertible and the IR background field is
homogeneous, $\phi(x)=\Phi$.

Now, each successive loop integral in the $n$-loop contributions which are
not explicitly written in Eq.~(\ref{loopexp}) brings a suppression factor
\begin{equation}
\frac{k^d-k^{\prime d}}{k^{\prime d}}=\ord\left(\frac{k-k'}{k'}\right)
\end{equation}
due to the integration volume in the momentum space. Thus
$\delta k/k=(k-k')/k$
appears as a new small parameter which suppresses the higher loop contributions
in the blocking relation and the ``exact'' functional differential equation
obtained in the limit $\delta k\to 0$ includes the one-loop contribution
only. But we should bear in mind that the loop expansion had to be
used at the initial stage of the derivation so the resulting
``exact'' equation might be unreliable in the strong coupling
situation. All we know is that the loop corrections to the
evolution equation obtained in the one loop level are
vanishing.

We will use the gradient expansion for the action,
\begin{equation}
S[\phi]=\sum_{n=0}^\infty \int \d^d x\, U_n(\phi(x),\partial^{2n}),
\end{equation}
where $U_n$ is an homogeneous function of order $2n$ in the derivative.
In the leading order of this expansion, the so-called
local potential approximation, we have
\begin{equation}
S[\phi]=\int\d^dx \left[\frac{Z(\phi)}{2}(\partial_\mu\phi)^2+U(\phi)\right],
\label{localpotapprox}
\end{equation}
and furthermore the simplification $Z(\phi)=1$ will be used to derive a
simple differential equation for the potential $U$. This
local potential will be then the only function characterizing the
action. If we use now a homogeneous infrared background field,
$\phi(x)=\Phi$, we obtain from Eqs.~(\ref{loopexp}) and~(\ref{localpotapprox})
an equation for the
local potential $U_k(\phi=\Phi)$:
\begin{equation}
U_{k-\delta k}(\Phi)=U_k(\Phi)+\frac{1}{2}\mathrm{tr}
\log[\square+U_k^{\prime\prime}(\Phi)]+\ord(\delta k^2),
\label{WHeq}
\end{equation}
where we have introduced the notation
\begin{equation}
U_k^{\prime\prime}(\Phi)=\frac{\partial^2 U_k(\Phi)}{\partial\Phi^2},
\quad \square=-\partial_\mu\partial^\mu
\end{equation}
and the trace is taken in the subspace of the eliminated modes.
We can explicitly write the trace in momentum space and get
\begin{equation}
U_{k-\delta k}(\Phi)=U_k(\Phi)+\frac{1}{2}\int
\frac{\d^d p}{(2\pi)^d}\log[p^2+U_k^{\prime\prime}(\Phi)]+\ord(\delta k^2),
\end{equation}
where the integration extends over the shell $k-\delta k<p<k$. In the limit
$\delta k\to 0$ one then finds the differential equation
\begin{equation}
k\frac{\partial}{\partial k}U_k(\Phi)=-\frac{\Omega_d k^d}{2(2\pi)^d}
\log[k^2+U_k^{\prime\prime}(\Phi)],
\label{WHeqcont}
\end{equation}
where $\Omega_d$ denotes the $d$-dimensional solid angle
\begin{equation}\label{solidc}
\Omega_d=\frac{2\pi^{d/2}}{\Gamma\left(\frac{d}{2}\right)}.
\end{equation}

The Wegner-Houghton equation~(\ref{WHeqcont})
represents the one-loop resummed
mixing of the coupling constants of the potential $U_k=\sum_n (g_n/n!)\Phi^n$.
In fact, an expansion of the logarithm in the second derivative of the
potential gives
\begin{equation}
k\frac{\partial}{\partial k}U_k(\Phi)=
-\frac{\Omega_d k^d}{2(2\pi)^d}\sum_{n=1}^\infty\frac{1}{n}
\left(\frac{-U_k^{\prime\prime}(\Phi)}{k^2+U_k^{\prime\prime}(\Phi)}
\right)^n,
\end{equation}
up to a field independent constant. This is the usual one loop resummation
of the effective potential~\cite{coleman73} except that the loop
momentum is now restricted to the subspace of the modes to be eliminated.
Actually, the fact that the r.h.s. includes the running potential
$U_k(\Phi)$ rather than the bare one, $U_\Lambda(\Phi)$, indicates that the
contributions of the successive eliminations of the degrees of freedom
are piled up during the integration of the differential equation and the
solution of the renormalization group equation resums the perturbation
series. The solution of the differential equation interpolates between the
bare and the effective potential as $k$ is lowered
from the original cutoff $\Lambda$ to zero.

Finally, let us note that the derivation of Eq.~(\ref{WHeqcont}) shows that
the restoring force for the fluctuations into the equilibrium is
proportional to the argument of the logarithm function. Thus a
nontrivial saddle point should be used when
\begin{equation}
k^2+U_k^{\prime\prime}(\Phi)\le0.
\label{broken}
\end{equation}

\section{Evolution of the coupling constants}
Our effective action is
\begin{equation}
S_k=\int \d^d x \left[\frac{1}{2}(\partial_\mu\phi(x))^2+
U_k(\phi(x))\right],
\end{equation}
and the initial condition for the evolution equation is given
at $k=\Lambda$. The potential $U_k(\Phi)$ is assumed to be
polynomial so it is expanded as
\begin{equation}
U_k(\Phi)=\sum_{n=0}^N \frac{1}{n!}g_n(\Phi_0)(\Phi-\Phi_0)^n.
\label{expansion}
\end{equation}
We study the model in the symmetric phase, where the saddle point is trivial,
$\Phi_0=0$. The polynomial structure of the potential
is consistent because we avoid the singularity of Eq.~(\ref{WHeqcont}) at
$k_\mathrm{cr}^2=-U_{k_\mathrm{cr}}^{\prime\prime}(\Phi)$
(recall Eq.~(\ref{broken})) which occurs in a region around $\Phi=0$ in the
symmetry broken phase~\cite{polonyi97}.

Taking the $n$-th derivative of Eq.~(\ref{expansion}) at $\Phi=0$
we obtain the
coupling constant $g_n(k)$, and define the corresponding beta function by
\begin{eqnarray}
g_n(k)&=&\frac{\partial^n}{\partial\Phi^n}U_k(\Phi)_{|\Phi=0},\nonu
\beta_n&=&k\frac{\d}{\d k}g_n(k)=
\frac{\partial^n}{\partial\Phi^n}k\frac{\partial}{\partial k}U_k(\Phi).
\end{eqnarray}
By taking the successive derivatives of Eq.~(\ref{WHeqcont}), we obtain
\begin{equation}
\beta_n=-\frac{\Omega_d k^d}{2(2\pi)^d}\mathcal{P}_n(G_2,\ldots,G_{n+2}),
\label{betafunct}
\end{equation}
where
\begin{equation}
G_n=\frac{g_n}{k^2+g_2}
\label{bigg}
\end{equation}
and
\begin{equation}
\mathcal{P}_n=\frac{\partial^n}{\partial\Phi^n}
\log[k^2+U_k^{\prime\prime}(\Phi)]
\end{equation}
is a polynom of order $n$ in the variables $G_j$, $j=2,\ldots,n+2$,
\begin{eqnarray}\label{polinomia}
\mathcal{P}_1&=&G_3, \nonumber \\
\mathcal{P}_2&=&G_4-G_3^2, \nonumber \\
\mathcal{P}_3&=&G_5-3G_3G_4+2G_3^3, \nonumber \\
\mathcal{P}_4&=&G_6-4G_5G_3-3G_4^2+12G_3^2G_4-6G_3^4,\\
\mathcal{P}_5&=&G_7-5G_6G_3-10G_5G_4+20G_5G_3^2+30G_4^2G_3-60G_4G_3^3+24G_3^5,
\nonumber \\
\mathcal{P}_6&=&G_8-6G_7G_3-15G_6G_4-10G_5^2+30G_6G_3^2+
120G_5G_4G_3+30G_4^3\nonu
&&-120G_5G_3^3-270G_4^2G_3^2+360G_4G_3^4-120G_3^6.\nonumber
\end{eqnarray}

The coupling constants defined through Eq.~(\ref{expansion})
are dimensional parameters (the field variable $\phi$ has dimension
$(d-2)/2$). However, the corresponding dimensionless parameters have more
physical sense. We obtain them in the following way:
\begin{equation}
\tilde g_n(k)=k^{-n(1-\frac{d}{2})-d} g_n(k)=
(\Lambda\tilde{k})^{-n(1-\frac{d}{2})-d} g_n(k),
\label{dimensionless}
\end{equation}
where now $\tilde{k}$ runs from 1 to 0. Their beta functions are
\begin{equation}
\tilde\beta_n=-\left[n\left(1-\frac{d}{2}\right)+d\right]\tilde g_n
+k^{-n(1-\frac{d}{2})-d}\beta_n,
\label{bettil}
\end{equation}
where the first and the second term stands for the tree-level
and the loop corrections, respectively. One can see that the
super-renormalizable coupling constants follow
asymptotically free scaling law at the tree level.

\section{UV scaling laws and their extensions}

One can distinguish an ultraviolet
and an infrared scaling regime, for $k^2\gg |m^2(k)|$ and for
$k^2\ll |m^2(k)|$, respectively.
In the UV regime the scale dependence comes dominantly from the
$k^2$ term of the propagator, see the denominator of Eq.~(\ref{bigg});
the $k$-dependence is generated by the phase factor $k^d$
in the IR regime where $k^2$ could be neglected in the inverse propagator.
We will begin at the UV scale with the
usual $\phi^4$ potential ($g_2\equiv m^2$)
\begin{equation}
V_\Lambda(\phi)=\hf m^2\phi^2+\frac{1}{4!}g_4\phi^4+\frac{1}{6!}g_6\phi^6,
\end{equation}
and see how the different couplings are generated when we move towards the
IR regime.

One ignores the $g_2$ term in the denominator of Eq.~(\ref{bigg})
in the asymptotic UV regime and finds
\begin{eqnarray}\label{betafo}
\frac{\d g_2}{\d k}&=&-\frac{\Omega_d}{2(2\pi)^d}\,k^{d-3}g_4,\nonu
\frac{\d g_4}{\d k}&=&\frac{\Omega_d}{2(2\pi)^d}\,k^{d-3}\left(
\frac{3}{k^2}g_4^2-g_6\right), \label{evolutions} \\
\frac{\d g_6}{\d k}&=&-\frac{\Omega_d}{2(2\pi)^d}\,3g_4k^{d-5}\left(
\frac{10}{k^2}g_4^2-5g_6\right),\nonumber
\end{eqnarray}
where in the last equation we omitted the contribution of $g_8$.
Consider the usual strategy in which the coupling constants $g_n$ are
neglected for $n>4$ and the resulting equation
is easy to integrate,
\begin{equation}
\frac{1}{g_4(k)}=\frac{1}{g_4(\Lambda)}+\frac{3\Omega_d
\left(1-\left(\frac{k}{\Lambda}\right)^{4-d}\right)}{(4-d)2(2\pi)^dk^{4-d}}.
\label{g4UVsc}
\end{equation}
This expression agrees with the result of the minimal
subtraction (MS), a scheme which proved to be specially
convenient in the ultraviolet scaling regime. It is based on the
analytical continuation of the loop integrals in the ultraviolet domain
so the resulting beta functions are mass independent, i.e.
the terms $\ord(g_2/k^2)$ are neglected. When extrapolating to the infrared
regime we find erroneously the mass independent result $g_4\sim k^{4-d}$
($g_4\sim \log k$ in $d=4$), $g_4$ tends to zero as $k\to 0$.
This can be understood by inspecting  (\ref{betafo}) where we find
large positive values in the infrared for $g_4\not=0$ (this
conclusion remains valid for finite $g_6$, as well). When the
mass term is retained the beta function assumes the correct behavior
and becomes $\ord(k^d)$ in the infrared.
Note that the term with $g_6$ acts in the opposite manner
than $g_4$, c.f. the different signs in the right hand side of
(\ref{betafo}), so that it can change the evolution considerably.

It is instructive to look into the evolution of the dimensionless
coupling constant,
\begin{equation}
\frac{1}{\tilde g_4(k)}=\frac{\left(\frac{k}{\Lambda}\right)^{4-d}}
{\tilde g_4(\Lambda)}
+\frac{3\Omega_d
\left(1-\left(\frac{k}{\Lambda}\right)^{4-d}\right)}{(4-d)2(2\pi)^d}.
\end{equation}
For $4-d>0$ the one-loop $\omega=4-d$ universal critical exponent
is reached for $k$ values sufficiently below the cutoff 
where $k/\Lambda\approx0$. The latter condition
is needed to get rid of the non-universal cutoff effects. The scaling
changes qualitatively as $d\to4$ because the non-universal
$k\approx\Lambda$ behavior is spread over the whole $k$ range
due to the smallness of $4-d$. This is what happens in the expansion
\be
1-\left(\frac{k}{\Lambda}\right)^{4-d}\to(d-4)\ln\frac{k}{\Lambda},
\end{equation}
employed in the dimensional regularization scheme. This generalizes
to any dimension: the marginal coupling constant follows the
scaling law extended from the non-universal cutoff regime.

The evolution for $g_2$ is of the form
\begin{equation}
\frac{\d g_2}{\d k}=-\frac{\Omega_d}{2(2\pi)^d}\,g_4 k^{d-3}.
\label{gtwoas}
\end{equation}
It predicts
\begin{equation}
\frac{\d g_2}{\d k}\sim k,~~~g_2\sim k^2+\text{const}
\end{equation}
in lack of any dimensional constant.

In the IR scaling regime we neglect the $k^2$ term in the denominator of
Eq.~(\ref{bigg}) and using Eqs.~(\ref{betafunct}) and~(\ref{polinomia}),
we get for $g_4$ (assuming again $g_6=0$),
\begin{equation}
\frac{\d g_4}{\d k}=\frac{3\Omega_d}{2(2\pi)^d}\,k^{d-1}\,\frac{g_4^2}{g_2^2}.
\label{g4IRsc}
\end{equation}
This evolution is much slower comparing with Eq.~(\ref{g4UVsc}).
In fact, for $d>1$ we have now a suppression factor $k^{d-1}$
which makes the coupling to stabilize at the attractive IR fixed point.
In the same way, we obtain for $g_2$
\begin{equation}
\frac{\d g_2}{\d k}=-\frac{\Omega_d}{2(2\pi)^d}\,k^{d-1}\,\frac{g_4}{g_2},
\label{g2IRsc}
\end{equation}
with a variation which is slower than that predicted by the UV
scaling and a suppression factor for $k\to 0$.

We see that the extraction of the scaling in the limit
$k\to 0$ from the UV scaling laws, which is the commonly accepted
practice in perturbation theory is incorrect when the MS scheme
is used. One has to come back to the complete scaling
laws in order to describe correctly the IR scaling.

Four-dimensional asymptotically free gauge models present an infrared
Landau pole in perturbation theory. But this behavior results from
the extrapolation from the UV scaling laws. As we have just
remarked, the IR limit of the UV regime is not correct in
general, and the mass term can change considerably the actual behavior
in the IR. Moreover, at the IR side of the UV regime, there are nonlinear
effects that make important the contribution of the
irrelevant (non-renormalizable) couplings (see Fig.~3 (b),
commented in the next Section),
therefore even the IR limit of the UV scaling can be influenced by these
couplings. These ideas have been considered qualitatively in the previous
paragraphs, after inspection of Eq.~(\ref{evolutions}). Let us now
examine them more quantitatively.

We take as an example of asymptotically free model the scalar theory in
three dimensions. We know from the epsilon-expansion result~\cite{wilson72}
that below four dimensions, the $\lambda\phi^4$ theory does not
present an infrared Landau pole, but a fixed point located at
\begin{equation}
\tilde{\lambda}^*=\frac{16\pi^2}{3}\epsilon,
\label{WFfixedpoint}
\end{equation}
at order $\epsilon=4-d$. This is the Wilson-Fisher fixed point. For
finite $\epsilon$, for example in two and three dimensions, we can get this
one-loop result from our beta functions of the dimensionless coupling constants
obtained from Eqs.~(\ref{bettil}) and~(\ref{betafunct}) in the asymptotic
UV regime (that is, ignoring $g_2$), which for the
$\tilde{g}_4$ coupling give
\begin{equation}
\tilde{\beta}_4=-(4-d)\tilde g_4+ 3\tilde g_4^2\frac{\Omega_d}{2(2\pi)^d},
\end{equation}
giving the IR fixed point
\begin{equation}
\tilde g_4^*=\frac{(4-d)2(2\pi)^d}{3\Omega_d}.
\end{equation}
Restricting ourselves to the $d=3$ case, we find
\begin{equation}
\tilde g_4^*=\frac{4\pi^2}{3}\simeq 13.15947.
\label{g4FP}
\end{equation}
However, in three dimensions $g_6$ is a marginal coupling, and it can be
generated by the RG flow, modifying the position of the
fixed point~(\ref{g4FP}). If we include $\tilde g_6$ in our analysis, we
get the beta functions
\begin{eqnarray}
\tilde{\beta}_4&=&-\tilde g_4+\frac{3}{4\pi^2}\tilde g_4^2-
\frac{1}{4\pi^2}\tilde g_6 \, ,\nonumber \\
\tilde{\beta}_6&=&\frac{-3}{4\pi^2}\tilde g_4\,(10 \tilde g_4^2-5\tilde g_6).
\end{eqnarray}
It is immediate to see that the zeros of these beta functions are at the point
\begin{eqnarray}
\tilde g_4^*&=&4\pi^2\simeq 39.4784, \nonumber \\
\tilde g_6^*&=&32\pi^4\simeq 3117.091 .
\label{fixpe}
\end{eqnarray}
But now let us consider the inclusion of a non-renormalizable coupling, $g_8$.
The beta functions for the $(g_4,g_6,g_8)$ model are
\begin{eqnarray}
\tilde{\beta}_4&=&-\tilde g_4+\frac{3}{4\pi^2}\tilde g_4^2-
\frac{1}{4\pi^2}\tilde g_6 \, ,\nonumber \\
\tilde{\beta}_6&=&\frac{-3}{4\pi^2}\tilde g_4\,(10 \tilde g_4^2-5\tilde g_6)
-\frac{1}{4\pi^2}\tilde g_8 \, , \label{betag8} \\
\tilde{\beta}_8&=&\tilde g_8-\frac{7}{4\pi^2}\,
(-90\tilde g_4^4+60\tilde g_4^2\tilde g_6-5\tilde g_6^2-4\tilde g_4\tilde g_8).
\nonumber
\end{eqnarray}
From them, one obtains the fixed point
\begin{eqnarray}\label{fixpk}
\tilde g_4^*=\frac{4\pi^2}{210}\left(195 +
\sqrt{29625}\right) \simeq 69.01, \nonumber \\
\tilde g_6^*=-4\pi^2 \tilde g_4^* + 3\tilde g_4^{*2}\simeq 11563, 
\label{fixed_point_g8} \\
\tilde g_8^*=15\tilde g_4^{*3}-60\pi\tilde g_4^{*2}\simeq 2.11\cdot 10^6.
\nonumber
\end{eqnarray}
These values are also obtained in the numerical integration of
Eqs.~(\ref{betag8}), independently of the initial values for the different
couplings (if they are different from zero, which corresponds to the Gaussian
fixed point). 

To assess the importance of this result, the difference between the
physics around the fixed points (\ref{fixpe}) and (\ref{fixpk}), 
recall that the modification
of the irrelevant operator set at the cutoff influences the overall scale
of the model. Thus one has to consider dimensionless quantities
in comparing the two coupling constant regions. The most obvious 
candidate, the dimensionless ratio between the mass and the four point
vertex, $g_2/g^2_4$, is trivially vanishing in our approximation.
But $g_6$ is dimensionless and its variation at the fixed points
indicates that no adjustment of the overall scale could bring the physics
of these two fixed points together.

As we have expected, the non-renormalizable coupling $g_8$
modifies the position of the fixed point without changing the blocking
procedure, turning to a situation of strong coupling dynamics in the IR.
This is because the linearity which one assumes to ignore the irrelevant
coupling constants is no longer valid in the strong coupling regime.
As $g_4$ and $g_6$ approach their large IR fixed point the linearization
fails and new scaling laws are found which in turn generate new relevant
operators \cite{polonyi97}, overlapping with $\phi^8$. Thus the
strong coupling dynamics may induce a new (and artificial) IR scaling
regime even if the UV scaling laws are extrapolated down to low energies.

Nothing unusual happens for infinitesimal
$\epsilon$ when one stays in the vicinity of the Gaussian fixed point.
In that case, linearity applies all along the RG flow from the Gaussian
fixed point to the Wilson-Fisher fixed point given by 
Eq.~(\ref{WFfixedpoint}). However, as we have seen, its location is
changed for finite $\epsilon$ by
nonlinear effects produced in the flow from one fixed point to
the other. In the vicinity of this infrared fixed point (which, 
we stress again, is artificial in the sense that it neglects the
influence of the mass term) we of course have again the classification
of relevant and irrelevant terms, which, in the case of three and
two dimensions, is different from the one obtained from power
counting. For example, this IR fixed point has in two dimensions
just the mass and the fourth order coupling as relevant parameters,
while higher order couplings are irrelevant.

It is well known that the poles of the fixed point
action at complex values of the field variable make the
Taylor expansion in the field unreliable \cite{trpot}. We do 
not see any reason to reject a blocked action only because the
potential is diverging beyond a given field strength. This kind of
internal space singularity might only indicate a
maximal particle density in the system. We should stay
only sufficiently far from this limiting value of the field variable
when the evolution equation is truncated. We interpret the
difference of the two fixed points as an indication of the breakdown
of the simple universality which is based on the linearized flow
equation around the UV fixed point.

So far we considered the extrapolation of the UV scaling laws to the
IR regime. Does the conclusion concerning the importance of the
non-renormalizable coupling constant $g_8$ remains valid when the true
evolution equation, with $g_2\not=0$, is considered? The mass slows down
the evolution but this may happen ``too late'' and the strong coupling
effects can be found on the true renormalization group trajectory for
small enough renormalized mass, close enough to the critical point.
When the mass is large then crossover freezes the evolution of the
coupling constants ``earlier'' and the linearization remains valid.
To demonstrate this case recall
that the IR limit of the UV regime means a fixed point for the
dimensionless couplings. For a super-renormalizable coupling constant
such as $g_4$, which
has positive dimension, this would mean that the dimensional coupling goes to
zero when $k \to 0$. However, we know that for a relevant coupling, the
dimensionless quantity diverges when $k\to 0$, so that the dimensional coupling
will take a finite value at the IR.
This reasoning can be explicitly checked in
Fig.~1, in which we consider the $d=3$ scalar theory with
just one coupling, $g_4$. The white points follow the evolution of the
UV regime and its extrapolation to $k=0$. We observe that indeed the
dimensionless coupling reaches the fixed point given by Eq.~(\ref{g4FP}),
while the dimensional coupling goes to zero. However, if one considers the
complete beta function, i.e. retaining $g_2$ (black points),
the behavior is the same in the UV regime,
but then the true trajectory separates
from the IR limit of this regime and enters into the actual IR regime, which
implies a divergent dimensionless coupling at $k=0$,
and a certain finite value of the
dimensional coupling, as explained above. One can however see numerically
that this finite value is stable and almost does not change when one
introduces more and more non-renormalizable couplings in the RG evolution.
This is what one expects when the crossover captures the coupling
constants and slows down their evolution in the regime of linearizability
where the non-renormalizable couplings are unimportant.

In the same way, it might
well be that the IR Landau pole
observed in four-dimensional gauge theories is just an artifact of
a wrong IR limit or the truncation of the renormalized action.
First, the nature of the singularity can change
when one adds non-renormalizable couplings, and then the IR Landau
pole would be the reflection of the insufficient functional form of the
blocked action, and second, the true IR trajectory
can be quite different from the IR limit of the UV scaling.

\section{Asymptotic freedom and the perturbation expansion}

We examine in this Section the scaling laws in dimensions
$d=2,3$ and 4 from the point of view of the
applicability of the perturbation expansion.

As we have seen in the previous Section, in the RG 
evolution of our model there are two asymptotic scaling regimes, 
$k\to\infty$ and $k\to0$. The latter one is trivial
as mentioned above, because the beta functions
\eq{betafunct} are suppressed by the factor $k^d$
and the evolution of the dimensional coupling constants
slows down as $k\to0$. The asymptotic UV scaling is 
however more involved.
The super-renormalizable (relevant) and
the renormalizable (marginal) coupling constants, $g_n$ with
$n<2d/(d-2)$ and $n=2d/(d-2)$ according to the power counting, 
respectively,
follow their autonomous evolution, the universal
renormalized trajectory\footnote{Ignoring the triviality in $d=4$
where the tree level marginal
coupling constant $g_4$ is actually irrelevant due to the
radiative corrections.}.
The non-renormalizable (irrelevant) coupling constants
``forget quickly'' their initial value and take values which
are generated by the universal flow. This general trend
is demonstrated by the renormalization group flow
shown for $d=4$ and 3 in Figs.~2 and~3,
respectively. In those Figures the dimensionless coupling constants 
are displayed
as the functions of the cutoff
which is measured in the units of the
initial cutoff value, $k\to k\Lambda$.
 
In the case of theories 
with non-Gaussian asymptotically free couplings 
(the case of our model for $d<4$), an excessive growth of
these couplings in the UV regime may produce a non-perturbative
situation in the infrared. We want to study this by comparing the
values of the couplings at $k=0$. To do so, 
we will adopt the convention that a model with positive
renormalized mass square is non-perturbative in the vertex
$g_n\phi^n$ if the radiative correction $\ord(g_n)$
to the self energy is stronger than the mass term, i.e.
\begin{equation}
\frac{\tilde g_n(k)}{((n-2)/2)! \, 2^{(n-2)/2}\tilde m^2(k)}\gg1
\label{pert_cond}
\end{equation}
where $\tilde m^2(k)=\tilde g_2(k)$.
In case this inequality were satisfied in the infrared
($k=0$), 
this would mean a non-perturbative situation and the invalidity of 
renormalized perturbation expansion.

We remark that we are 
asking here about the validity of the perturbative condition 
at the true, $k=0$, 
infrared fixed point, where the RG flow ends.
We are not considering for example the situation at 
the Wilson-Fisher IR fixed point
in $d<4$, which, as was explained in Section 4, 
is the IR limit of the UV scaling behavior (a fixed point
which describes the behavior of the system at the end of 
the UV regime), and not the 
real IR fixed point if we ask for the behavior of the
system at energy 
scales much lower than the mass $m(k)$ (for example, in
dimension two, we will see that at the
crossover, or the end of the UV regime, the high order
couplings start to take large values, while they are
irrelevant couplings for the Wilson-Fisher IR fixed
point; this is because the true RG flow separates from
the extrapolation of the UV flow, as was explicitely shown
in Fig. 1 in the case of $d=3$).

We now turn to a detailed analysis of the
situation at dimensions four, three and two.
Since the neglected higher order
vertices may influence the evolution while we lower the cutoff,
we have to address the problem of the
system of coupled equations numerically.

The evolution of a non-renormalizable coupling constant, $\tilde g_6$ in
four dimensions, is shown in Fig.~2. The irrelevance is
expressed by the independence of $\tilde g_6(k)$ on the initial value
$\tilde g_6(1)$ for $k\ll1$\footnote{The different initial value
for the non-renormalizable
coupling constants may induce a different overall scale factor.
This effect is very weak in our case due to the smallness of the
renormalizable coupling constants.}. The value given in the
leading order of the perturbation expansion,\footnote{Which is not 
applicable for the strongly coupled case~\eq{fixpe}
or~\eq{fixpk}.} $\ord(\tilde g_4^3)$,
is reached at $k\approx0.3$ for not too large values
of $\tilde g_6(1)$. Since the model is in the weak coupling regime
the evolution is rather slow after arriving at this universal
value if the cutoff is high enough to provide a long scaling
regime. In our case the scale window $0.3<k<1$ was insufficient
and the plateau is reduced into a peak at $k\approx0.3$ before the
crossover. But the bringing of $\tilde g_6(1)$
close to the universal
value creates a plateau even with this limited range of the scales
as it can be seen for $\tilde g_6(1)=9\cdot10^{-8}$. Such a
scale independence is the ultimate goal of the improved action
program~\cite{impr}.
As the initial value increases beyond the
plateau level the coupling constant decreases in a monotonous manner
as the cutoff is lowered. In order to check the critical exponent
coming from the linearized blocking we need a $\tilde g_6(k)$
larger than the universal value since the latter originates in the
nonlinear level, $\ord(\tilde g_4^3)$.
The evolution follows the
linear relation $\tilde g_6\approx k^2$
for $k<1$ indicated by the dashed line in the
last plot of Fig.~2. The infrared scaling regime is
$0<k<0.1$ where $\tilde g_6(k)$ tends to zero with $k$ along the universal
trajectory. The theory remains perturbative in $d=4$ since it has no
relevant non-Gaussian coupling constant.

The three-dimensional renormalization group flow is depicted in Fig.~3. 
The asymptotic infrared scaling laws are rather
simple, the super-renormal\-iz\-able coupling constants diverge,
the renormalizable one $n=6$ converges and the non-renormalizable
ones tend to zero in the infrared, $k\to0$. The ultraviolet
scaling law, above $k\approx0.1$ indicates the weak, radiative correction
generated relevance of $\tilde g_6$ for the given initial conditions,
$\tilde g_n(1)$. The insensitivity on the initial
condition $\tilde g_8(1)$ for $k<0.3$ seen in the last plot supports
the irrelevance of $\tilde g_8$. In fact, the evolution of
$\tilde g_8(k)$ follows the linearized scaling law as long as
its value is far form the universal $\ord(\tilde g_4^4)$ value.
It is worthwhile noting that the non-renormalizable
coupling constants $\tilde g_n(k)$ always develop a peak of sign
$(-1)^{1+n/2}$ around the crossover, $k\approx0.1$. The appearance of the
peak can be understood as the result of the increase of
$|\tilde g_n(k)|$ from zero as the cutoff is lowered in the ultraviolet
scaling regime and the decrease in the infrared side of the crossover.

Non-perturbative phenomena may arise at the low energy
edge of the UV scaling regime due to the increase of the
asymptotically free coupling constants,
$g_4$ and $g_6$ if the scaling regime is long
enough and the initial value of the coupling constants
$\tilde g_4(1)$ and $\tilde g_6(1)$ are large enough.
There is however no problem in finding a perturbative, 
asymptotically free theory in the infrared.
The parameter with the highest energy dimension in the Lagrangian
is $g_2=m^2$ for $d>2$. Thus $\tilde g_2$
is the largest among the dimensionless coupling constants
according to Eq.~\eq{bettil}, and it dominates the
action and renders the theory perturbative in the IR limit.
The comparison of this conclusion with Eq.~\eq{fixed_point_g8}
reveals the necessity of treating the IR scaling laws properly in 
establishing the validity of the renormalized perturbation
expansion. 

As $d$ approaches 2 more and more coupling constants become
super-renormalizable. The fastest increasing dimensionless
non-Gaussian coupling constant during the decrease of the
cutoff is $\tilde g_4$. The critical exponent, the measure of
the speed of the increase become degenerate for infinitely
many coupling constants when $d=2$. The specialty of the
lower critical dimension is the existence of infinitely
many super-renormalizable coupling constants, $\tilde g_n$,
with equal critical dimension. This degeneracy of the
dimensions evolves the non-Gaussian pieces of the action
with the same rate as the mass term on the tree level
and the theories are not obviously perturbative any more.
In other words, it remains for the radiative corrections,
the last term in the right hand side of Eq.~\eq{bettil}
to determine if the theory runs into weak- or
strong-coupling regime at the infrared. 

The renormalization group equations were integrated out numerically
in two dimensions
with the initial conditions $(g_2^i=-0.001, g_4^i=0.01, g_n^i=0, n>4)$ to
find the evolution of the coupling constants. The result,
depicted in Fig.~4,
shows a marked increase at the low energy end
of the UV scaling regime. This increase originates from
the asymptotically free evolution. The relevant
behavior of a coupling constant is defined on the linearized
level of the blocking, i.e. in the leading order of the
perturbation expansion. This one-loop result can be
obtained by replacing the running coupling constants
in the beta functions by their initial values at the cutoff.
The local potential obtained in the one-loop approximation is
\begin{eqnarray}
U_k(\phi)&=&V_\Lambda(\phi)+\frac{1}{2}\int_{k<p<\Lambda}
\frac{\d^2 p}{(2\pi)^2}
\log[p^2+V_\Lambda^{\prime\prime}(\phi)]= \nonumber \\
&&\frac{1}{8\pi}\left\{[\Lambda^2+V_\Lambda^{\prime\prime}(\phi)]
\log[\Lambda^2+V_\Lambda^{\prime\prime}(\phi)]- \right. \nonumber \\
&&\left.[k^2+V_\Lambda^{\prime\prime}(\phi)]
\log[k^2+V_\Lambda^{\prime\prime}(\phi)]-\Lambda^2+k^2\right\},
\label{oneloop}
\end{eqnarray}
with
\begin{equation}
V_\Lambda^{\prime\prime}(\phi)=m^2(\Lambda)+\frac{g_4(\Lambda)}{2}\phi^2.
\end{equation}
The comparison of the numerical solution with the one-loop
evolution is shown in Fig.~5.
The one-loop formula~(\ref{oneloop}) cannot be extended down
to $k=0$, because
there will be a value of $k$ such that $k^2+m^2(\Lambda)=0$, since
$m^2(\Lambda)$ is negative. But we only want to compare the results in
the perturbative regime. So we have stopped the evolution in
Fig.~5 at $k\sim 0.3$. One can also see an increase in the
one-loop solution at small $k$ which accumulates and drives
the system non-perturbative at lower values of $k$.

The numerical results of Fig.~4 show that the
initial conditions $(g_2^i=-0.001, g_4^i=0.01, g_n^i=0, n>4)$
correspond to a non-perturbative system. Can we find
initial conditions which yield perturbative dynamics?
In order to answer this question the left hand side
of the inequality \eq{pert_cond} is plotted against
the initial value for $g_2$ on Fig.~6,
for the different couplings up to $N=20$,
at a value of $g_4^i=0.001$. The result does not
change qualitatively for different values of bare $g_4$.
It supports the general trend of having the systems more
perturbative when the Gaussian part of the action is
increased. The higher order
coupling constants tend to grow faster but it seems that $g_n$
can be brought into the perturbative regime for sufficiently
large initial mass square. At $g_2^i=0.01$, for example, all
the couplings are perturbative, and this perturbative character is
more pronounced for the high couplings. However, the separation between the
values of the l.h.s. ratio of Eq.~\eq{pert_cond} at
$g_2^i=0.01$ is smaller
as we go to higher couplings, and one can ask whether it has got a
limiting value or the trend can be reversed for a sufficiently
high order coupling. Indeed, Fig.~7 reveals that
this happens for the $n=24$ coupling ($N=26$),
which suggests that one will
have a non-perturbative situation also at this value of
bare $g_2$ going to a sufficiently high order coupling constant.
The situation is the same, even stronger, below two dimensions:
the existence of an infinite number of relevant couplings makes
that one cannot assure perturbativity by looking to a finite
number of couplings, no matter what the initial conditions are.
We take this and similar other failures in finding a
perturbative theory observed at different initial conditions as a
strong numerical indication of the non-perturbative nature of
{\em any} two and lower dimensional scalar field theory with
polynomial interaction.

\section{Lattice Wegner-Houghton equation}
\label{lattice}
The previous Sections dealt with the renormalization group flow
at finite scales. We address now a different, asymptotic problem, 
the manner the sensitivity on the initial values
of the irrelevant coupling constants is suppressed 
during the renormalization. 
This question is usually rendered trivial
by the universality argument. But there are two reasons to
suspect that such a reasoning which is based on the linearization
of the blocking relation might be oversimplified; both are 
related to an infinite set of operators.

The reason motivating a more careful check of the
universality, mentioned in the Introduction, is that the 
models at or below the lower critical dimension contain 
infinitely many relevant operators. It is not obvious whether
the sum over the interaction vertices is always convergent
enough to make the linearization of the blocking relation
a reliable approximation.

Another potential problem shows up if one changes infinitely
many irrelevant terms in the action by choosing another regulator.
Let us compare the momentum space cutoff in the continuum with
the lattice regularization. The propagator is a monotonic
function of the momentum in the continuum. This is not the case
on the lattice. In fact,
the fermion doubling problem on the lattice~\cite{nogo} results
from the periodicity of the propagator in the first Brioulline zone,
the appearance of $2^d-1$ new maxima in the propagators in the UV,
non-universal regime. The existence of a maximum of the
propagator in the UV regime contradicts an assumption 
of the studies of the continuum models, namely that the propagator
decreases monotonically as $p\to\infty$, and renders the perturbation
expansion non-universal for lattice fermionic models~\cite{reisz}. 
There is no species doubling for bosons but
their propagator remains periodic on the lattice and we find
$2^d-1$ lattice extrema in the UV regime. The existence of these 
extrema is an evidence of the slowing down in the decrease of the
propagator as the momentum approaches the boundary of the first 
Brioullin zone. This in turn indicates the weaker suppression 
of the high energy modes compared to the continuum regularization. 
Does this mean that the UV scaling laws are different in the
continuum than on the lattice? We shall find an affirmative
answer to this question but this result does not contradict
the universality.

Some words of caution are in order at this point. One would object the
interpretation of the modification of the cutoff as the introduction
of new irrelevant coupling constants by recalling that the theory
ceases to be renormalizable in the presence of the non-renormalizable
(irrelevant)
couplings. The resolution of the apparent paradox is based on the
difference between the ways
the renormalization group is used in Statistical
and High Energy Physics. We are interested in the dynamics close
to the cutoff in Statistical Physics and this is respected
by the employment of the blocking which keeps the {\em complete}
dynamics unchanged below the actual cutoff. The price of this
precision is the appearance of the infinitely many irrelevant
coupling constants in the action. We seek the dynamics at
finite, fixed scales in High Energy Physics. Since the cutoff is
sent to the infinity this boils down the problem of keeping
the physics cutoff independent {\em far from the cutoff} only.
The obvious gain of such an ease of the conditions is the
freedom from the adjustment of the non-renormalizable
parameters. Thus one can remove the cutoff when the non-renormalizable
parameters are present in the action without any problem\footnote{Ignoring
again the possibility of the triviality,
the appearance of an UV Landau pole.} as long as the
renormalization conditions are imposed far from the cutoff.

The lattice regularization of the scalar model can be described
by using the momentum space as the introduction of the non-renormalizable
higher order derivative terms,
\bea
(\partial_\mu\phi)^2&\longrightarrow&
\frac{4}{a^2}(\sin\frac{a\partial_\mu}{2i}\phi)^2\\
&=&\left(\sum\limits_{\ell=0}^\infty
\frac{1}{(2\ell+1)!}\left(\frac{a}{2}\right)^{2\ell}\partial^{2\ell+1}
\phi\right)^2.\nonumber
\eea
The cutoff dependence of the non-renormalizable coupling constants
follows a tree-level relation arising from the Taylor expansion
of the sine function. This is sufficient to establish convergent
physics at finite scales when $a\to0$~\cite{reisz}, a claim to
be verified in this Section numerically by means of the implementation of
the Wegner-Houghton scheme on the lattice. But this convergence
can not rule out a modification of the scaling laws in the
asymptotical UV regime. In fact, we shall find a new scaling regime
between the region where the usual universal UV scaling is observed
and the UV fixed point. The only effect the different
adjustments of the non-renormalized coupling constants may leave on the
finite scale physics can be comprised in an overall scale factor.

It is rather straightforward to repeat the steps leading to
Eq.~\eq{WHeqcont} on the lattice. It is shown in the Appendix
that the only change required is the modification of the ``solid angle''
factor, $\Omega(k)$: the lattice evolution equation
\eq{WHeqlatt} is obtained from Eq.~\eq{WHeqcont} by replacing
Eq.~\eq{solidc} by Eq.~(\ref{omega3}). One recovers the continuum solid
angle for $d=2$, $\Omega_2=2\pi$, in Eq.~(\ref{omega3}) as $k\to0$,
thus the WH equations agree in the IR limit.
In fact, one sees numerically that the behavior of the evolution of the
different coupling constants in the lattice RG is qualitatively the same
as in the continuum case. But the question we are interested in
is the relation between the regularizations in the UV, where
the coupling constants are introduced, when the physics is the same
at finite scales. It is shown in the Appendix that there is a
natural relation between the cutoffs, $\Lambda^2=8/a^2$, which
matches the finite scale physics. We shall follow the renormalization
group flow in terms of the coupling constants whose dimension is removed
by the initial value of the cutoff,
\begin{equation}
g_n \longrightarrow g_n/\Lambda^2,
\label{rescaling}
\end{equation}
in order to avoid the singularities at $k=0$.

Let us consider $\lambda\phi^4$ lattice theory which can be studied
either numerically or analytically and whose properties can be matched
to those of the continuum theory by the adjustment of $g_2$ and $g_4$.
But the situation is more involved in two dimensions. The reason is again
that there are infinitely many renormalizable coupling constants
and one cannot match the finite scale physics by adjusting $g_2$ and $g_4$.
This is demonstrated in Fig.~8 where the lattice
model with the initial conditions $m^2(8)=g_2(8)=-0.001$, $g_4(8)=0.01$,
and $g_n(8)=0,~n>4$ (where we have already made the rescaling
Eq.~(\ref{rescaling})) was evolved in the infrared direction. As the system
reached $k=k_\mathrm{end}$ the continuum
WH equation was used to increase the cutoff. The result is a
``perfect matching'' of the models in the UV which gives the same low
energy physics in the IR. As we can see in Fig.~8, the lattice
$\lambda\phi^4$ model in two dimensions is {\em not} the continuum
$\lambda\phi^4$ theory. It contains contributions of infinitely many
renormalizable other coupling constants.
Of course, numerically we had to truncate the equations
at a certain coupling (here, at $\ord(\Phi^{22})$) but we checked that these
``truncation effect'' hardly influences the values of the low order coupling
constants.

We have taken for the parameter $k^2_\mathrm{end}$
the value $k^2=0.3$ in Fig.~8, while the crossover
is at $k^2_\mathrm{cr}\sim 0.01$.
We had to use $k^2_\mathrm{end}>k^2_\mathrm{cr}$,
because the high order couplings have very large values at the crossover
which requires very fine discretization in the numerical
resolution of our differential equations to ensure that the way back to the
UV is done accurately. However, $k^2_\mathrm{end}$ should also be
sufficiently small that the flow be universal there, in other words
to make sure that the irrelevant lattice contributions are suppressed
for $k^2<k^2_\mathrm{cr}$.

The choice $k^2_\mathrm{end}>0$ introduces an uncertainty in the
matching. To assess it we repeated the ``go-return'' evolution
described above and checked the discretization errors for
$k^2_\mathrm{end}=0.8,~0.5$ and $0.3$.
After then we took the appropriate bare parameters at the
UV end points in both regularizations and followed the evolutions
down to $k=0$. The relative difference,
\begin{equation}
\Delta g_n\equiv
\frac{|g_n^\mathrm{CONT}(0)-g_n^\mathrm{LATT}(0)|}{|g_n^\mathrm{LATT}(0)|}
\label{relative}
\end{equation}
is shown in Table~1 for the different coupling constants
at $k=0$. The smallness of the deviation assures that
the IR behavior has practically been obtained with $k^2_\mathrm{end}=0.3$,
and one can trust the conclusions, the approach of the flow to a
universal curve, extracted from Fig.~8.

The first two plots in Fig.~8 show that the mass and the
quartic coupling constant run parallel in the SUV region of the lattice
regularization (see Appendix) and in the continuum. There is no
convergence between the two regularizations in this unusual scaling regime,
anticipated above. The approach to the universal curve starts
for $k^2<4$, below the SUV regime only. The fact that the renormalization group
flow converges to the universal one in the $2^{-d}$-th part
of the Brioullin zone only sets an unexpected high lower limit on the
lattice size when the continuum limit is sought in numerical simulations.
The higher order vertices seem to converge to the universal curve from the
very beginning but the difference between the two regularizations is
surprisingly large. The universal trajectory of the $\phi^4$ model is
reached later by the higher order vertices. This effect appears to be
a counterpart of the non-perturbative features seen in
Figs.~6,~7 and introduces a large uncertainty in
identifying two-dimensional models in different regularizations.

\section{Conclusions}
The renormalization group flow of scalar models with
polynomial interaction is considered in the first part of this paper
by solving the Wegner-Houghton equation numerically in the local
potential approximation for $d=2,~3$ and 4. 

The numerical results showed in this paper suggest
that the length of the UV scaling regime which is
needed to generate non-perturbative dynamics in the infrared shrinks
to zero as the number of the asymptotically free coupling
constants tends to infinity. In other words, the
$\Lambda$-parameter tends to the cutoff as the
lower critical dimension is approached, $d\to2$.
Such a behavior limits considerably the values of the coupling
constants for a perturbative system in dimension 3
and renders {\em all} two- and lower dimensional field theories
with polynomial couplings non-perturbative.
This makes the understanding of the noncritical low
dimensional condensed matter systems more involved. 
The one-dimensional models
belong to first quantized quantum mechanics and our result is
a manifestation of the failure of the convergence of the
perturbation expansion for an anharmonic oscillator.

Such a conclusion does not invalidate the well known results
for two-dimensional systems, such as the applicability of the
Bethe ansatz, bosonization and the availability of certain exact
information for models with conformal invariance. Instead, it makes
the asymptotic state structure and the relation between the
the dressed particles and the states created by the application
of the field operator from the vacuum more involved.

We found an interesting analogy between the infrared
Landau pole of the confining four-dimensional Yang-Mills theories
and the low dimensional scalar models which opens the possibility of
an unexpected, nontrivial structure in the asymptotic states in
the low dimensional scalar models. Viewed with interest in particle 
physics our conclusion suggests that one can 
avoid the IR Landau pole by following the evolution of the
non-renormalizable operators. 

{\em How to find the non-renormalizable operators whose presence 
stabilises the theories at low energies?} 
It is well known that massive Lagrangians generate
trivial infrared scaling laws, i.e. the Gaussian mass term is
the only relevant operator in the infrared scaling regime.
This is because the fluctuations are exponentially
suppressed beyond the correlation length so the evolution
of the coupling constants slows down at the infrared side
of the crossover. The theories with dimensional
transmutation, i.e. dynamically generated scale parameter
or infrared instability only can support non-perturbative
dynamics in the IR scaling regime. Thus the operators
sought should be relevant in the IR regime, their growth
being fed by IR or collinear divergences. There are few
known cases only where the low modes are controlled
by non-renormalizable operators. These include
the four fermion contact term in solids inducing the BCS
transition~\cite{shankar}, the higher order derivative
terms in the action which generate inhomogeneous vacuua~\cite{afvac}, 
the common element being the onset of a Bose-Einstein
condensation~\cite{polonyi97}.

In the second part of the paper the infinitesimal renormalization
group scheme is generalized for lattice regularization. The matching of
the continuum and the lattice regularizations is carried out
numerically and the approach of the universal renormalization group flow
is demonstrated for the two-dimensional $\phi^4$ lattice model.
This result suggests that the naive argument for the universality,
which is based on the linearization of the blocking relations
remains valid in the presence of infinitely many relevant
operators. Other potential troublemakers, the infinitely 
many higher order derivatives contained in the lattice 
kinetic energy do generate a new,
``super UV'' scaling regime but universality is restored
at the IR end of the usual UV scaling regime. Another
use of the lattice regulated version of the Wegner-Haughton equation 
is the estimate of the finite size effects in a non-perturbative manner.
This provides a useful check of the thermodynamic limit of the
numerical results obtained in general on small lattices.

\section*{Appendix}
The details of the derivation of the Wegner-Houghton
equation in the lattice regularization are given in this Appendix.
Let us consider the scalar field theory
regularized on a lattice of lattice spacing $a=1$.
We want to derive a WH equation similar to Eq.~(\ref{WHeqcont}).
We integrate over spherical shells in the momentum space for the
continuum regularization, because the propagator has spherical symmetry.
This is no longer the case on the lattice, where
we have
\begin{equation}
\square=\sum_{\mu=1}^d\hat p_\mu^2, \quad
\hat p_\mu=2\sin\frac{p_\mu}{2}.
\end{equation}
Let us see the surfaces of equal value of the lattice propagator
in two dimensions by performing the following change of variables:
\begin{equation}
\renewcommand{\arraystretch}{1.5}
\left.
\begin{array}{c}
(p_x,p_y)\longrightarrow (p,\theta) \\
4\sin^2\frac{p_x}{2}+4\sin^2\frac{p_y}{2}=p^2 \\
\tan\theta=\frac{\sin (p_y/2)}{\sin (p_x/2)}
\end{array}
\right\}
\label{chvar}
\end{equation}
We can see in Fig.~9 the form of the curves of constant propagator
for several values of $p^2$. $p$ can be identified
as the ``momentum scale'' that runs from the cutoff at $p^2=\Lambda^2=8$ to
the IR $p^2=0$. It is also clear in that figure that the value $p^2=4$
separates two regimes, still in the ultraviolet region, that we could call
super-UV (SUV), for $8>p^2>4$, and normal-UV regimes. For $p^2\sim 0$,
the lines are spheres, and our change of variables~(\ref{chvar}) reduces to
the usual relation between cartesian and polar coordinates.

The absolute value of the Jacobian of the transformation~(\ref{chvar})
is found to be
\begin{equation}
J=\frac{J_\mathrm{p}}{\sqrt{\left(1-\frac{p^2}{4}\cos^2\theta\right)
\left(1-\frac{p^2}{4}\sin^2\theta\right)}},
\end{equation}
where $J_\mathrm{p}$ is the usual Jacobian for the polar change of
variables, $J_\mathrm{p}=p$. The transformation~(\ref{chvar}) can be
easily generalized to three and four dimensions; however, we can only
treat analytically the integral that appears in the derivation of the
WH equation in the $d=2$ case.

To derive the equivalent of the Wegner-Houghton equation~(\ref{WHeqcont})
in the bidimensional lattice regularization we will start from
Eq.~(\ref{WHeq}), and calculate the trace by integrating in momentum space
over a shell $k-\delta k<p<k$, where $p$ is the parameter we have introduced
in Eq.~(\ref{chvar}),
\begin{eqnarray}
\frac{1}{2}\mathrm{tr}\log[\square+U_k^{\prime\prime}]&=&
\frac{1}{2}\int\frac{\d^2 p}{(2\pi)^2}\log[4\sin^2\frac{p_x}{2}+
4\sin^2\frac{p_y}{2}+U_k^{\prime\prime}]= \nonumber \\
&&\frac{1}{2(2\pi)^2}\int\d\theta\int_{k-\delta k}^k \d p\,\, J\,
\log[p^2+U_k^{\prime\prime}]\approx \nonumber \\
&&\frac{\delta k}{2(2\pi)^2} k \log[k^2+U_k^{\prime\prime}]\,\Omega(k),
\end{eqnarray}
with
\begin{equation}
\Omega(k)=\int\d\theta\, \frac{8}{\sqrt{64-16 k^2+k^4\sin^2 2\theta}}.
\label{omegageneral}
\end{equation}
We have to distinguish two different regimes in making the integration:
\medskip
(i) $\underline{k^2<4.}\quad$
In this region the range of values for $\theta$ is
$(0,2\pi)$.
\begin{equation}
\Omega(k)=\int_0^{2\pi}\d\theta\, \frac{8}{\sqrt{64-16 k^2+k^4\sin^2 2\theta}}=
\sum_{i=1}^4\Omega_i(k),
\end{equation}
where we have split the interval $(0,2\pi)$ into four intervals
$(0,\pi/2)$, $(\pi/2,\pi)$, etc. Let us consider $\Omega_1(k)$. With
the change of variable $x=\tan\theta$ and the notation
\begin{equation}
\bar k^2=4-k^2,
\end{equation}
this integral can be brought into the form
\begin{equation}
\Omega_1(\bar k)=\frac{1}{\bar k}\int_0^\infty
\frac{2\,\d x}{\sqrt{(x^2+b^2)(x^2+b^{-2})}},
\label{integral1}
\end{equation}
where $b^2=4/\bar k^2$. The integral in Eq.~(\ref{integral1}) is
related to the elliptic integral of the first kind~\cite{tablas80}
$F(\phi,t)$,
\begin{equation}
\Omega_1(k)=F\left[\frac{\pi}{2},\frac{k}{4}\sqrt{8-k^2}\right];\quad
F(\varphi,t)=\int_0^\varphi\frac{\d\alpha}{\sqrt{1-t^2\sin^2\alpha}}.
\end{equation}
The result is the same for the other integrals $\Omega_i$, $i=2,3,4$.
But $F(\pi/2,t)$ is a complete elliptic integral, which can be
expressed in terms of
the hypergeometric function~\cite{tablas80,tablas65}
\begin{equation}
F(\alpha,\beta;\gamma;z)=\frac{\Gamma(\gamma)}{\Gamma(\alpha)\Gamma(\beta)}
\sum_{n=0}^\infty\frac{\Gamma(\alpha+n)\Gamma(\beta+n)}{\Gamma(\gamma+n)}
\frac{z^n}{n!}
\label{hypfunct}
\end{equation}
as
\begin{equation}
\Omega(k)=2\pi F\left(\frac{1}{2},\frac{1}{2};1;\frac{(8-k^2)k^2}{16}\right).
\label{omega1}
\end{equation}

\medskip
(ii) $\underline{4<k^2<8.}\quad$
This is the SUV region. We split again
the integral into four integrations in the corresponding quadrants.
By using the same change of variables as above, we have to calculate
\begin{equation}
\Omega_i(\bar k)=\int\frac{4\,\d x}{\sqrt{x^2(\bar k^4+16)-4\bar k^2(1+x^4)}},
\end{equation}
where
\begin{equation}
\bar k^2=k^2-4.
\end{equation}
Special care is needed at the limits of integration (recall
Fig.~9). It can be seen that the four integrals can be put
together in the form
\begin{eqnarray}
\Omega(\bar k)&=&4\int_{\bar k/2}^{2/\bar k}
\frac{4 \,\d x}{\sqrt{x^2(\bar k^4+16)-4\bar k^2(1+x^4)}}=\nonumber \\
&&\frac{8}{\bar k}\int_{\bar k/2}^{2/\bar k}
\frac{\d x}{\sqrt{(x-\bar k/2)(2/\bar k-x)(x+\bar k/2)(x+2/\bar k)}}.
\end{eqnarray}
This integral is related again~\cite{tablas80} to an elliptic integral
and an hypergeometric function:
\begin{equation}
\Omega(k)=\frac{32}{k^2}F\left[\frac{\pi}{2},\frac{8}{k^2}-1\right]=
\frac{16\pi}{k^2}F\left(\frac{1}{2},\frac{1}{2};1,\frac{(8-k^2)^2}{k^4}\right).
\label{omega2}
\end{equation}

\medskip
We would like to have a common expression for $\Omega(k)$ for both cases
(i) and (ii). From Eqs.~(\ref{omega1}) and~(\ref{omega2}), we find in fact
that the expressions differ in a factor 2 for $k^2=4$! The reason is that
actually our integral is divergent at this point. From Eq.~(\ref{omegageneral})
we see that the divergent integral is
\begin{equation}
\Omega(k^2=4)=\int_0^{2\pi}\d\theta\, \frac{2}{|\sin 2\theta|}
\end{equation}
(in fact, the hypergeometric function~(\ref{hypfunct}) converges in general
only in the unit circle $|z|<1$~\cite{tablas80}).
We will see, however, that this divergence is
integrable during the RG evolution from $k^2=8$ to $k^2=0$,
and therefore it has no physical significance.
In order to have a consistent,
single expression for the cases $8<k^2<4$ and $4<k^2<0$, we will make use
of the following property of the hypergeometric functions~\cite{tablas80}:
\begin{equation}
F\left(\frac{1}{2},\frac{1}{2};1,z^2\right)=
\frac{1}{1+z}F\left(\frac{1}{2},\frac{1}{2};1,\frac{4z}{(1+z^2)^2}\right),
\quad 0\leq z<1.
\label{prop}
\end{equation}

Let us consider the expression~(\ref{omega1}) which is
valid for $k^2<4$. Using the property~(\ref{prop}) one finds,
\begin{equation}
\Omega(k)=2\pi F\left(\frac{1}{2},\frac{1}{2};1,
\frac{4z}{(1+z^2)^2}\right)=
2\pi(1+z)F\left(\frac{1}{2},\frac{1}{2};1;z^2\right),
\label{useofprop}
\end{equation}
where we have set
\begin{equation}
\frac{4z}{(1+z^2)^2}=\frac{(8-k^2)k^2}{16}.
\end{equation}
This equation has two solutions for $z$ as a function of $k$:
\begin{equation}
z=\quad \frac{k^2}{8-k^2}\quad; \quad \frac{8-k^2}{k^2},
\end{equation}
but only the first one is admissible for Eq.~(\ref{useofprop}), because
it gives $z<1$ for $k^2<4$, while the second solution gives a
value greater than 1 in this region.
Now, if we define $\tilde{k}=8-k^2$, then we have
$z=(8-\tilde{k}^2)/\tilde{k}^2$, and Eq.~(\ref{useofprop}) becomes
\begin{equation}
\Omega(\tilde{k})=2\pi\frac{8}{\tilde{k}^2}
F\left(\frac{1}{2},\frac{1}{2};1;z^2\right).
\end{equation}
Comparing this last expression with the result~(\ref{omega2}) for the
case $8<k^2<4$, we can finally write
\begin{eqnarray}
\Omega(k)&=&\frac{16\pi}{\bar k^2}
F\left(\frac{1}{2},\frac{1}{2};1;z^2\right), \text{ with }
z=\frac{8-\bar k^2}{\bar k^2} \nonumber \\
&\text{and}&\begin{cases}
  \bar k^2=k^2 & \text{if $k^2=8\ldots 4$},\\
  \bar k^2=8-k^2 & \text{if $k^2=4\ldots 0$}.
  \end{cases}
\label{omega3}
\end{eqnarray}

The hypergeometric function $F\left(\frac{1}{2},\frac{1}{2};1;z^2\right)$
can be computed directly from its definition~(\ref{hypfunct}). One can
obtain a high precision in the evaluation of the series with a reasonable
number of terms (say, around 50) when $z$ is not very close to 1, say,
for $0.7>z>0$. For $1>z>0.7$ we have used the following alternative
formula~\cite{tablas65}
\begin{eqnarray}
F(\alpha,\beta;\alpha+\beta;z)&=&
\frac{\Gamma(\alpha+\beta)}{(\Gamma(\alpha)\Gamma(\beta))^2}
\sum_{n=0}^\infty \frac{\Gamma(\alpha+n)\Gamma(\beta+n)}{(n!)^2}
[2\psi(n+1)\nonu
&&-\psi(\alpha+n)-\psi(\beta+n)-\log(1-z)](1-z)^n, \nonumber \\
&&(|\mathrm{arg}(1-z)|<\pi, |1-z|<1),
\end{eqnarray}
where~\cite{tablas65}
\begin{equation}
\psi(z)=\frac{\d \,\log\Gamma(z)}{\d z},
\end{equation}
which gives a better convergence for the function
$F\left(\frac{1}{2},\frac{1}{2};1;z^2\right)$ near $z=1$ because it
is a series in the variable $(1-z^2)$.

In conclusion, our generalization of the WH equation~(\ref{WHeqcont})
for a lattice regularization in two dimensions is
\begin{equation}
k\frac{\partial}{\partial k}U_k(\Phi)=
-\frac{\Omega(k) k^2}{2(2\pi)^2}\log[k^2+U_k^{\prime\prime}(\Phi)],
\label{WHeqlatt}
\end{equation}
where $k$ is the parameter $p$ of Eq.~(\ref{chvar}), and
$\Omega(k)$ is given by Eq.~(\ref{omega3}).

\section*{\protect Acknowledgements}

We wish to thank V. Branchina, S.B. Liao, J. Alexandre, H. Mohrbach,
E. Vicari and A. Pelissetto for useful discussions.
Work partially supported by the Spanish MEC, Acci\'on Integrada 
hispano-francesa HF1997-0041, the French program, Actions Integr\'ees
franco-espagnol, Picasso 98064. 
J.M.C. acknowledges support from the EU TMR program
ERBFMRX-CT97-0122. He also thanks the Spanish MEC, the CAI European
program and DGA (CONSI+D) for financial support.

\newpage
\bigskip
\centerline{\textbf{FIGURE CAPTIONS}}
\begin{description}
\item[Fig. 1:] Numerical RG evolution for the (a) dimensionless $\tilde g_4$
coupling and (b) dimensional $g_4$ coupling in the $d=3$
$g_4\phi^4$ scalar theory. Black points result from the integration of the
complete beta function with $g_2=0.052$,
while white points show just the UV regime (see text).
In (a), the IR limit of the UV regime is given by the
fixed point~(36).
\item[Fig. 2:] Renormalization group flow for the dimensionless coupling
constant $\tilde g_6(k)$ in $d=4$. The initial conditions are
$\tilde g_2(1)=\tilde g_4(1)=0.01$, together with the values of
$\tilde g_6(1)=0.0,~4\cdot10^{-8},~9\cdot10^{-8},~10^{-7}$
for the different lines in plot (a), and $\tilde g_6(1)=10^{-3}$ for plot (b).
In this last plot the points of the numerical renormalization group
flow are joined
by the linearized scaling $\tilde g_6\approx k^2$, shown by a dashed line.
\item[Fig. 3:] (a) Renormalization group flow for the dimensionless coupling
constants $\tilde g_n(k),$ $n=2,4,6$ and 8 in $d=3$.
The initial conditions are
$\tilde g_2(1)=\tilde g_4(1)=0.01$, $\tilde g_n(1)=0$ for $n\ge6$.
(b) Evolution of $\tilde g_8(k)$ with the same initial conditions
that before except for $\tilde g_8(1)=6\cdot10^{-5}$,
together with the linearized
scaling law $\tilde g_8(k)\approx k$, indicated by the dashed line.
\item[Fig. 4:] Renormalization group flow for the dimensional coupling
constants in $d=2$. The initial conditions are
$g_2(1)=g_4(1)=0.01$, $g_n(1)=0$ for $n\ge6$.
\item[Fig. 5:] Evolution of the dimensional coupling constants
obtained in the one-loop approximation (squares) and numerically (circles)
for $d=2$.
\item[Fig. 6:] The left hand side of Eq.~(41) as a function
of the bare mass square for $n=4,\cdots,20$ in $d=2$.
\item[Fig. 7:] The same as Fig.~6 but now including the $n=24$ coupling.
\item[Fig. 8:] Matching of lattice and continuum regularizations; we start with
lattice field theory at the UV, go down to the IR (circles) and then we
come back to the UV with the continuum regularization (squares); see text.
\item[Fig. 9:] Lines of equal value of the lattice propagator. To the inside, 
the lines drawn correspond to $p^2=7.9,7,6,5,4,3,2,0.5,0.1$.
\end{description}

\newpage
\bigskip
\centerline{\bf{TABLE CAPTIONS}}
\begin{description}
\item[Table 1:] Relative differences between the continuum and lattice
coupling constants at $k=0$, Eq.~(46), after having matched
the couplings at the UV using the parameter $k^2_\mathrm{end}$.
\end{description}

\newpage
\centerline{\bf TABLE 1}
\vspace{100pt}

\begin{table}[h]
\begin{center}
\begin{tabular}{cccc}\hline
& $k^2_\mathrm{end}=0.8$ & $k^2_\mathrm{end}=0.5$ & $k^2_\mathrm{end}=0.3$ \\
\hline \hline
$\Delta g_2$ & 0.0166 & 0.0096 & 0.0052 \\
$\Delta g_4$ & 0.0028 & 0.0011 & 0.0002 \\
$\Delta g_6$ & 0.0188 & 0.0116 & 0.0069 \\
$\Delta g_8$ & 0.0199 & 0.0135 & 0.0092 \\
$\Delta g_{10}$ & 0.1793 & 0.0992 & 0.0494 \\
$\Delta g_{12}$ & 0.0680 & 0.0424 & 0.0259 \\
$\Delta g_{14}$ & 0.0522 & 0.0340 & 0.0221 \\
$\Delta g_{16}$ & 0.9300 & 0.5558 & 0.3226 \\
$\Delta g_{18}$ & 0.2116 & 0.1261 & 0.0731 \\
$\Delta g_{20}$ & 0.0584 & 0.0389 & 0.0259 \\
\hline
\end{tabular}
\end{center}
\end{table}

\newpage
\centerline{\bf FIGURE 1}
\vspace{100pt}

\begin{figure}[h]
\centerline{\epsfig{figure=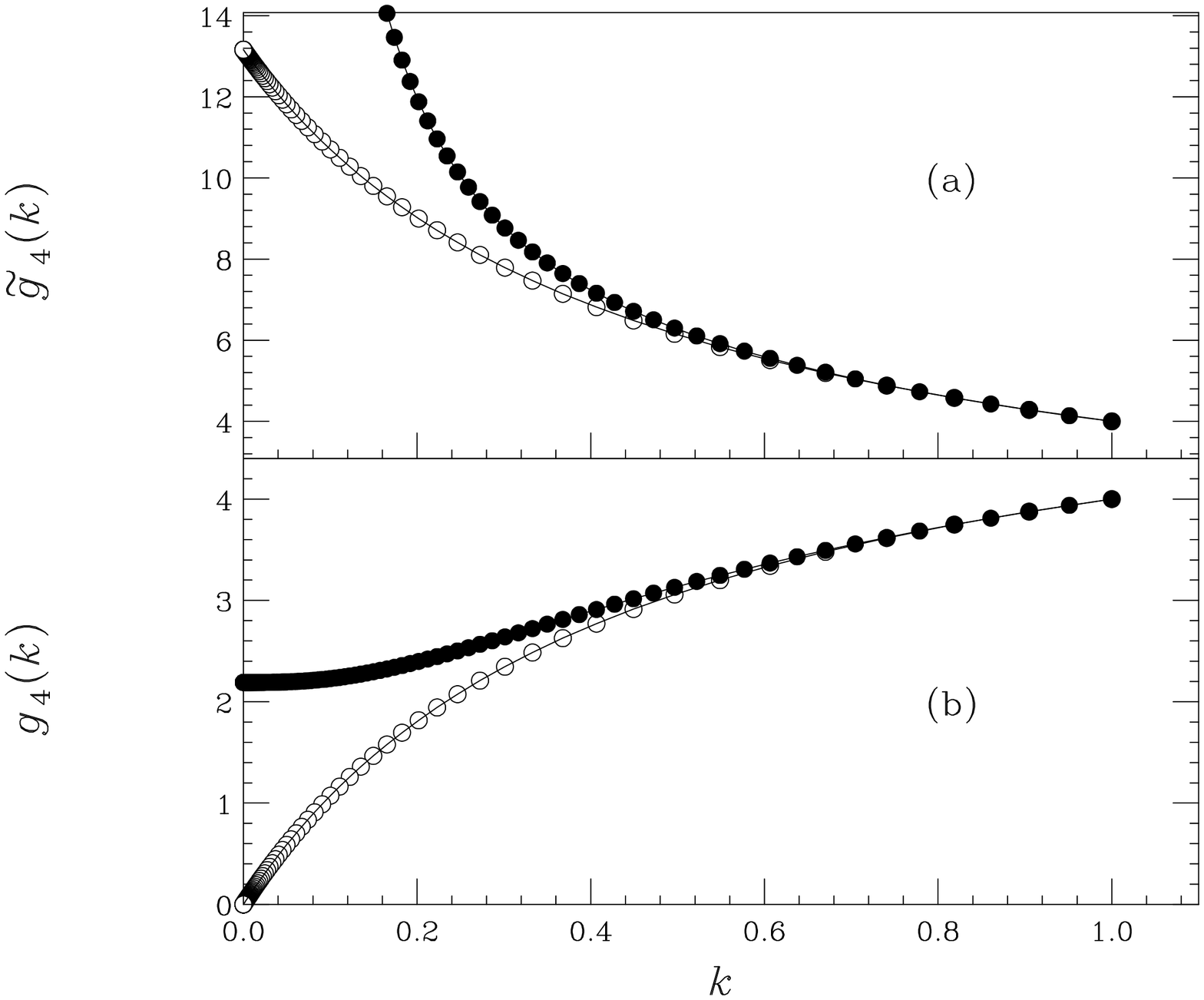,angle=90,width=170mm}}
\end{figure}
\newpage
\centerline{\bf FIGURE 2}
\vspace{100pt}

\begin{figure}[h]
\centerline{\epsfig{figure=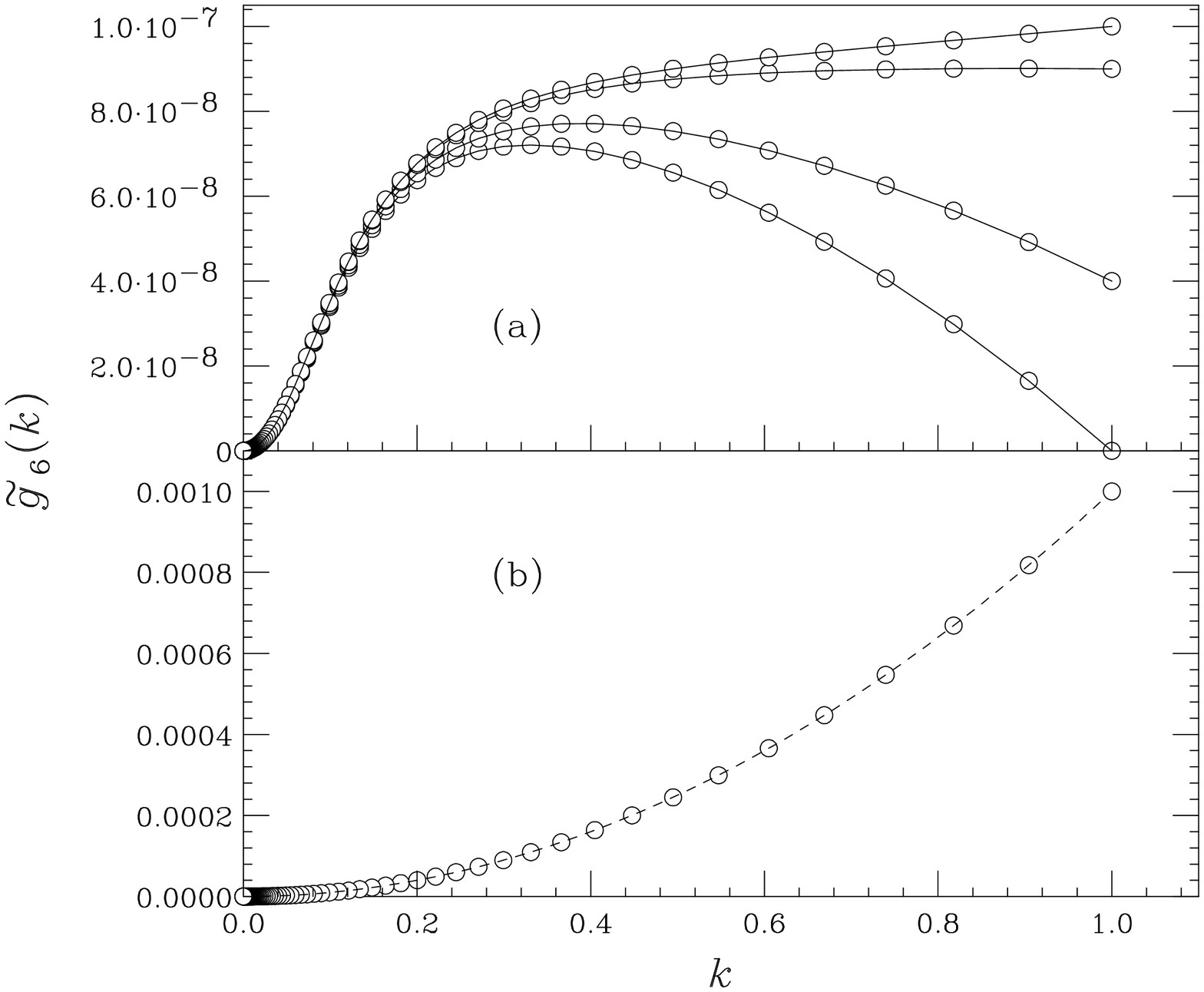,angle=90,width=170mm}}
\end{figure}

\newpage
\centerline{\bf FIGURE 3}
\vspace{100pt}

\begin{figure}[h]
\centerline{\epsfig{figure=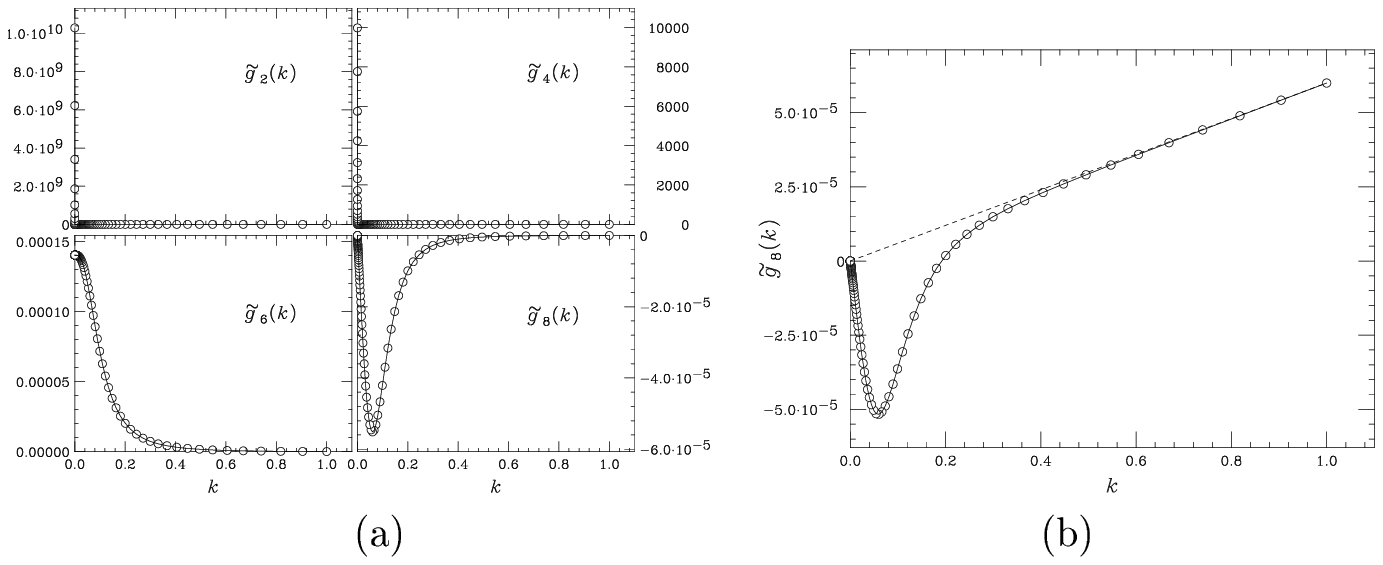,width=170mm}}
\end{figure}

\newpage
\centerline{\bf FIGURE 4}
\vspace{100pt}
\begin{figure}[h]
\centerline{\epsfig{figure=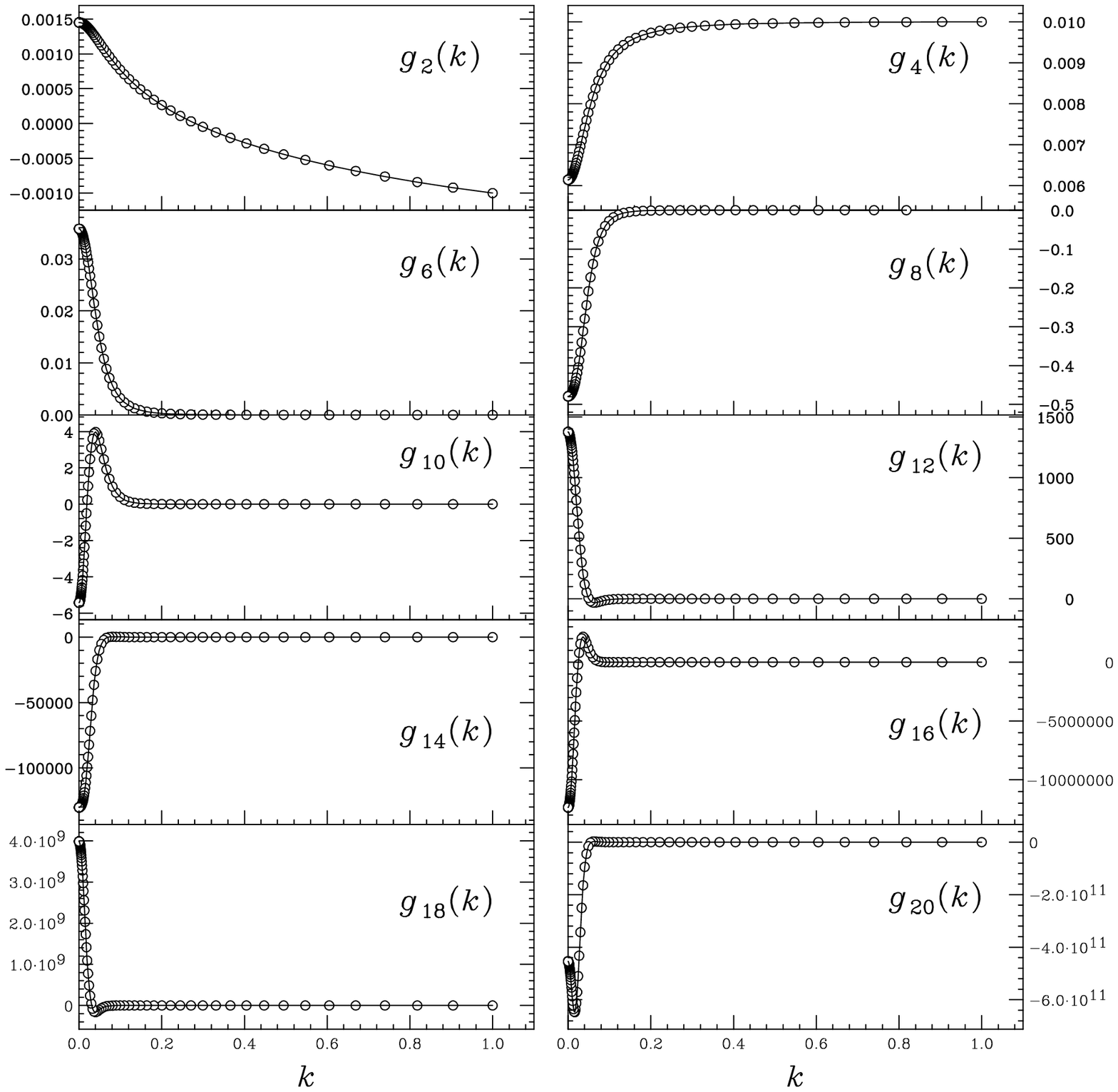,angle=90,width=150mm}}
\end{figure}

\newpage
\centerline{\bf FIGURE 5}
\vspace{100pt}

\begin{figure}[h]
\centerline{\epsfig{figure=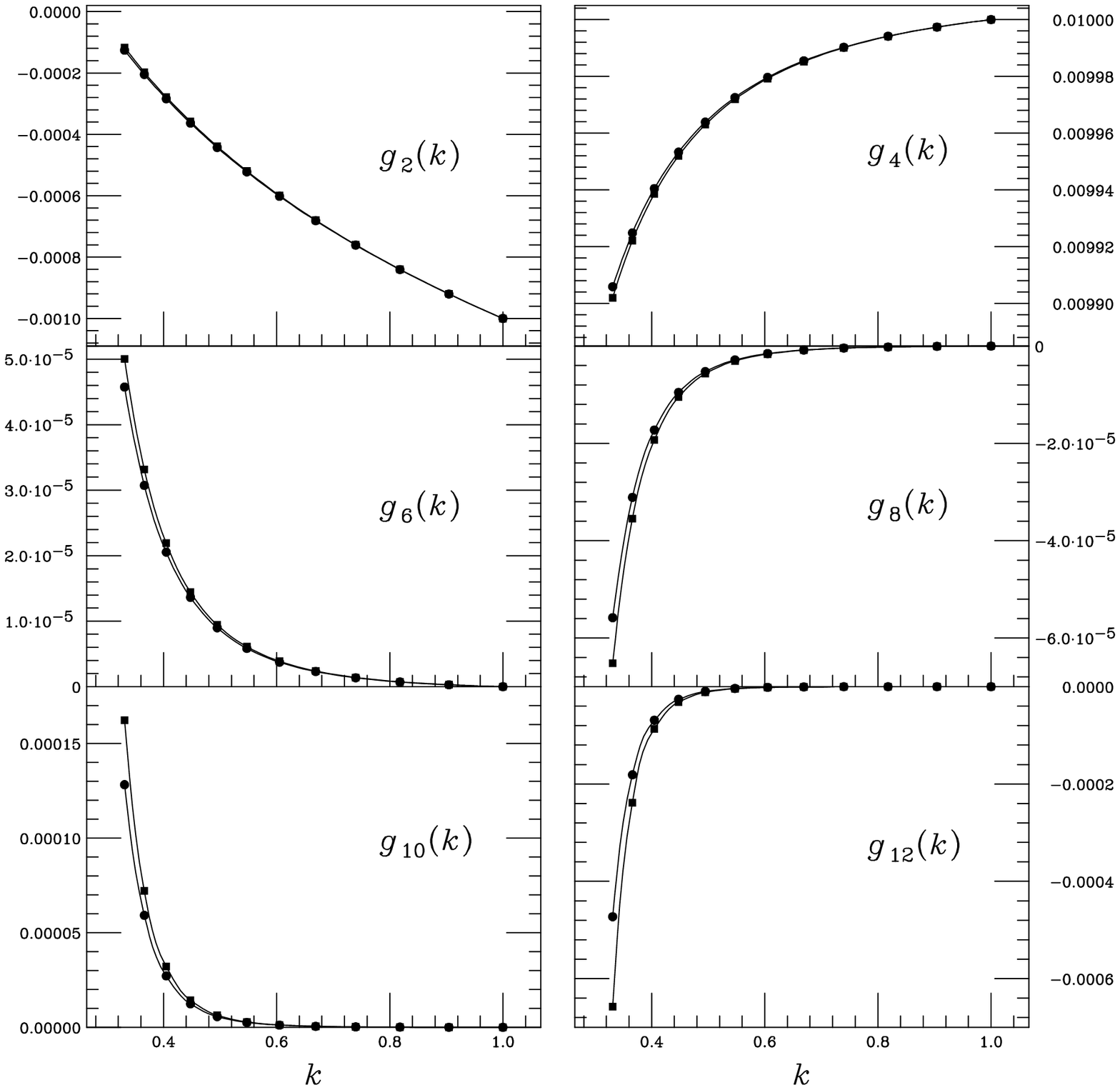,angle=90,width=150mm}}
\end{figure}

\newpage
\centerline{\bf FIGURE 6}
\vspace{100pt}

\begin{figure}[h]
\centerline{\epsfig{figure=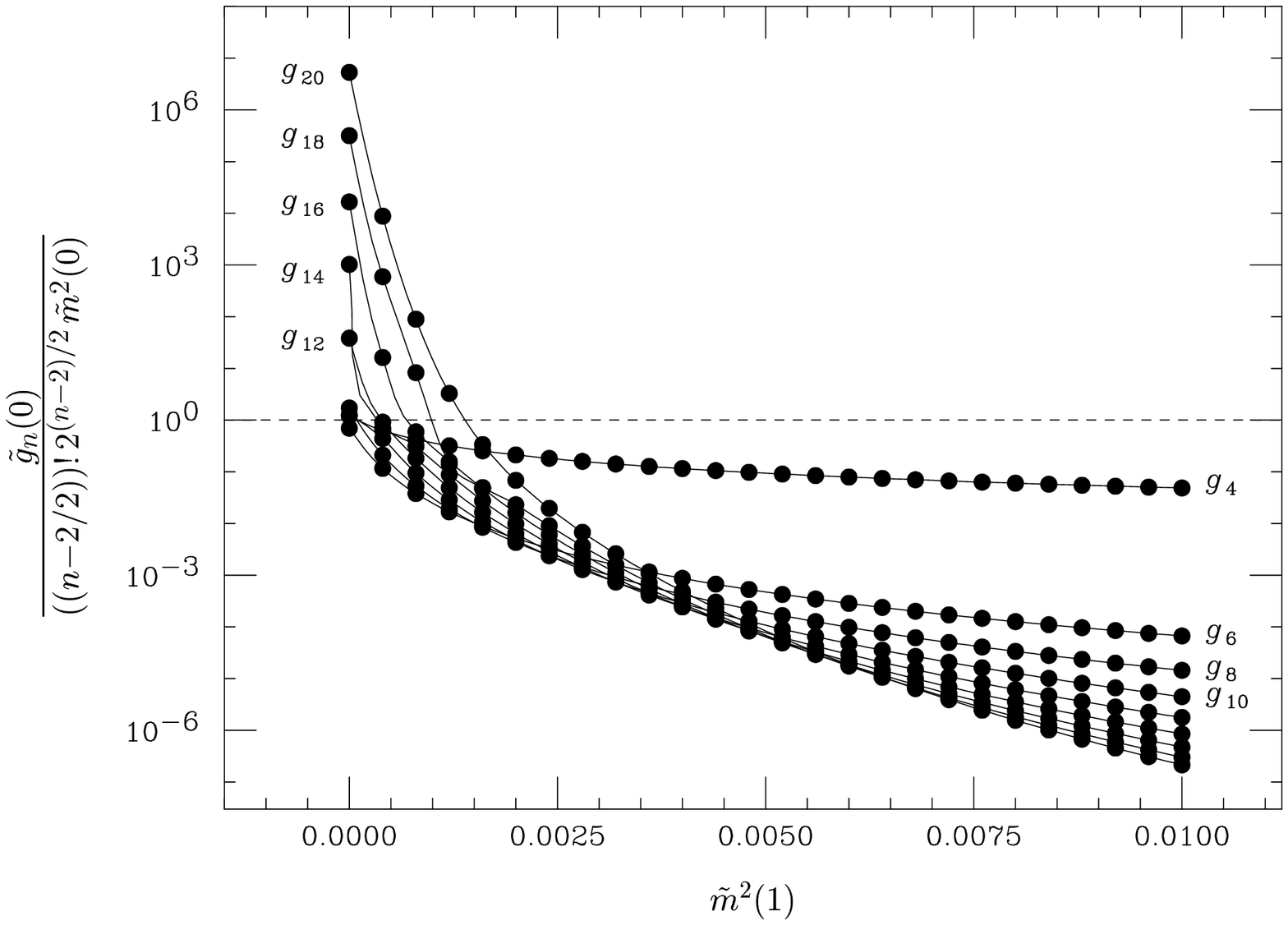,angle=90,width=170mm}}
\end{figure}

\newpage
\centerline{\bf FIGURE 7}
\vspace{100pt}
\begin{figure}[h]
\centerline{\epsfig{figure=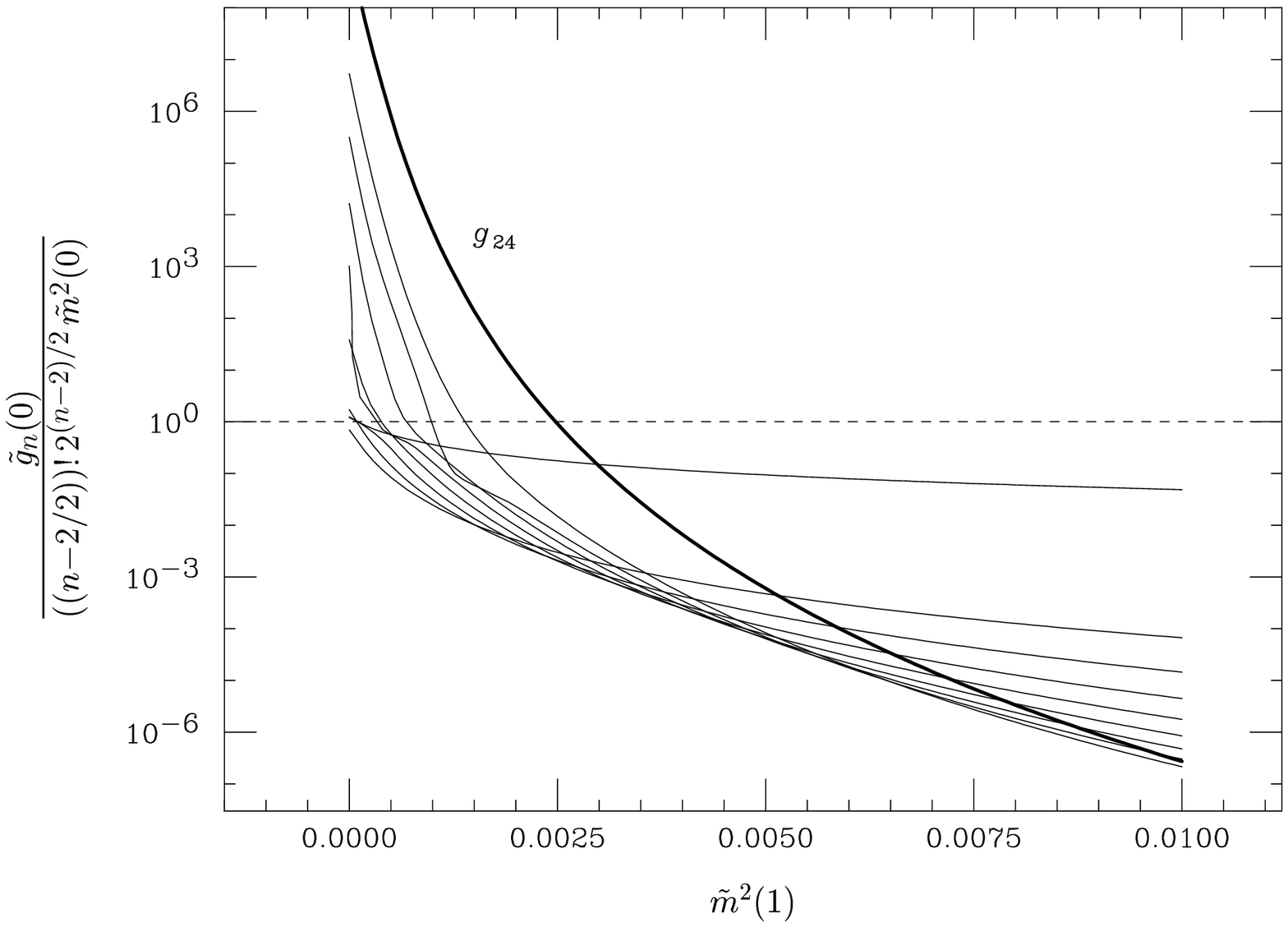,angle=90,width=170mm}}
\end{figure}

\newpage
\centerline{\bf FIGURE 8}
\vspace{100pt}

\begin{figure}[h]
\centerline{\epsfig{figure=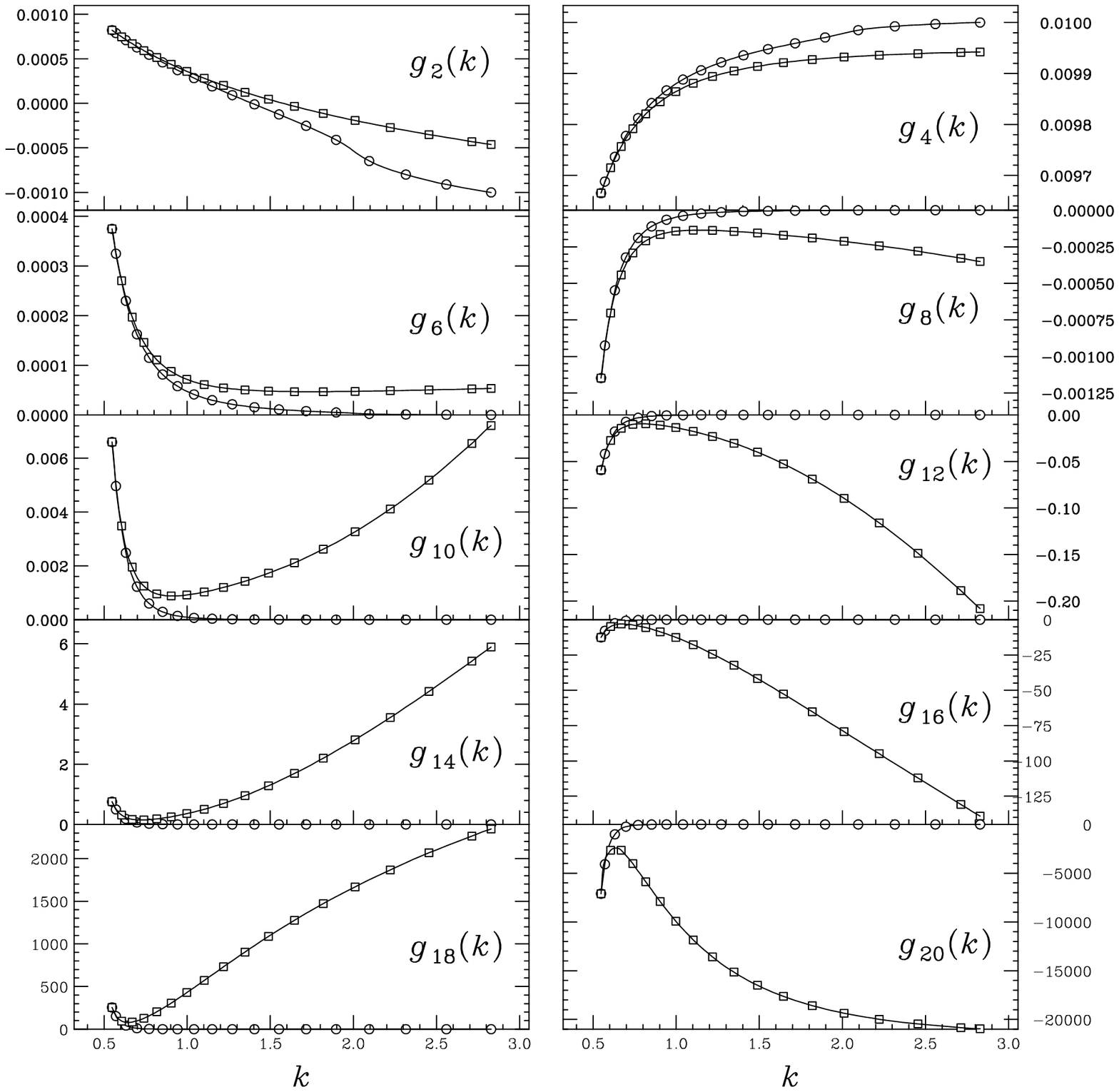,angle=90,width=150mm}}
\end{figure}

\newpage
\centerline{\bf FIGURE 9}
\vspace{100pt}

\begin{figure}[h]
\centerline{\epsfig{figure=fig9.ps,angle=0,width=150mm}}
\end{figure}

\end{document}